\RequirePackage[l2tabu, orthodox]{nag}
\RequirePackage{snapshot}

\documentclass[10pt,onecolumn]{extarticle}

\makeatletter
\if@twocolumn
  \usepackage[dvips,letterpaper,top=0.5in, bottom=0.5in, left=0.75in, right=0.5in,includefoot,heightrounded]{geometry}
\else
  \usepackage[dvips,letterpaper,margin=1in,includefoot,heightrounded]{geometry}
\fi

\usepackage{srcltx}

\usepackage[russian,portuges,english]{babel}

\iflanguage{portuges}
    {\newcommand{\keywordname}{Palavras-chaves}}
    {\newcommand{\keywordname}{Keywords}}

\usepackage{amsmath}
\usepackage{amssymb}
\usepackage{amsfonts}
\usepackage{mathtools}

\usepackage{abstract}

\usepackage{graphicx}
\usepackage[usenames,dvipsnames,svgnames,x11names]{xcolor}
\usepackage{subfigure}

\usepackage{booktabs}

\usepackage{setspace}
\usepackage{flushend}

\usepackage{multicol}

\usepackage{cite}

\usepackage{hyperref}
\usepackage[normalem]{ulem}

\usepackage{enumerate}

\usepackage[noend]{algpseudocode}

\usepackage{listings}

\usepackage[short,12hr]{datetime}

\usepackage{siunitx}

       \usepackage{fouriernc}

\usepackage{bm, bbm}
\usepackage{xfrac}
\usepackage{rotating}
\usepackage{ragged2e}
\usepackage{subfigure}
\usepackage{booktabs}

\def \E{\mathbbm{E}}

\def \No{\mathcal{N}}
\def \G{\mathcal{G}^0_{I}}

\newtheorem{COR}{Corollary}[section]
\newtheorem{LEM}{Lemma}[section]

\usepackage{natbib}
\setcitestyle{authoryear,open={((},close={))}} %

\newcommand{\printtitle}{%
\makeatletter
\if@twocolumn

\twocolumn[%
  \maketitle
  \begin{onecolabstract}
    \myabstract
  \end{onecolabstract}
  \begin{center}
    \small
    \textbf{\keywordname}
    \\\medskip
    \mykeywords
  \end{center}
  \bigskip
]
\saythanks
\else
  \maketitle
  \begin{onecolabstract}
    \myabstract
  \end{onecolabstract}
  \begin{center}
    \small
    \textbf{\keywordname}
    \\\medskip
    \mykeywords
  \end{center}
  \bigskip
  \onehalfspacing
\fi
\makeatother
}

\author{%
A.~D.~C.~Nascimento%
\thanks{
Departamento de Estat\'istica, Universidade Federal de Pernambuco,
Recife,
Brazil.
E-mail: \url{abraao@de.ufpe.br}
}
\and
J.~M.~Vasconcelos%
\thanks{
Departamento de Estat\'istica e Inform\'atica, Universidade Federal Rural de Pernambuco,
Recife,
Brazil.
E-mail: \url{josimar.vasconcelos@ufrpe.br}
}
\and
R.~J.~Cintra%
\thanks{%
Signal Processing Group,
Department of Technology,
Universidade Federal de Pernambuco,
Caruaru, Brazil.
E-mail: \url{rjdsc@de.ufpe.br}}
\and
A.~C.~Frery%
\thanks{
School of Mathematics and Statistics, Victoria University of Wellington,
Wellington, New Zealand.
E-mail: \url{alejandro.frery@vuw.ac.nz}
}
}

\title{%
Regression Model for Speckled Data with Extremely Variability}

\newcommand{\myabstract}{%
			Synthetic aperture radar (SAR) is an efficient and widely used remote sensing tool.
			However, data extracted from SAR images are contaminated with speckle, which precludes the application of techniques based on the assumption of additive and normally distributed noise.
			One of the most successful approaches to describing such data is the multiplicative model, where intensities can follow a variety of distributions with positive support.
			The $\mathcal{G}^0_I$ model is among the most successful ones.
			Although several estimation methods for the $\mathcal{G}^0_I$ parameters have been proposed, there is no work exploring a regression structure for this model.
			Such a structure could allow us to infer unobserved values from available ones.
			In this work, we propose a $\mathcal{G}^0_I$ regression model and use it to describe the influence of intensities from other polarimetric channels.
			We derive some theoretical properties for the new model: Fisher information matrix, residual measures, and influential tools.
			Maximum likelihood point and interval estimation methods are proposed and evaluated by Monte Carlo experiments.
			Results from simulated and actual data show that the new model can be helpful for SAR image analysis.
}

\newcommand{\mykeywords}{%
Speckled data,
			$\G$ regression model,
			influence measure,
			conditional intensity
}

\date{}

\begin{document}

\printtitle

	\section{Introduction}

	Synthetic Aperture Radar (SAR) sensors are powerful remote sensing devices.
	A SAR system transmits electromagnetic pulses toward a target and records the returned signal.
	Such a signal is rich in information because the physical properties of the environment modify it.
	Modeling the signal intensities is, therefore, a goal for SAR image understanding.

	SAR images are formed with coherent illumination and are therefore contaminated by an interference pattern called speckle~\citep{HypothesisTestingSpeckledDataStochasticDistances}.
	This pattern leads to deviations from additivity and Gaussian regularity assumptions, requiring tailored modeling and analysis techniques.
	Unlike other methods, this modeling has a phenomenological character closely related to image formation physics.

	\citet[][Sec.~5-7.1]{MicrowaveRadarRadiometricRemoteSensing} discuss a physical model for speckle formation.
	According to that model, if, per cell,
	\begin{enumerate}
		\item the scatterers' amplitude and phase are statistically independent,
		\item the phase carries no information, i.e., it is uniformly distributed in $[-\pi,\pi)$,
		\item no scatterer dominates the return, and
		\item there are infinitely many scatterers,
	\end{enumerate}
	then the multilook intensity return follows a gamma distribution, and this situation is called ``fully-developed speckle.''

	In practice, one or more of those hypotheses may not be valid.
	In particular, when the resolution cell is small relative to the targets size, e.g., over urban areas, or when the scatterers' spatial organization is not random, for example when observing forests on undulated relief, they do not hold and the intensity is seldom gamma-distributed.

	The multiplicative model (MM) is one of the most successful approaches to describe speckled data~\citep{SARImageStatisticalModelingPartISinglePixelStatisticalModels,SARImageStatisticalModelingPartIISpatialCorrelationModelsandSimulation} even under departures from those hypotheses.
	A special case of MM is the $\G$-distribution~\citep{freryetal1997a}, which depends on three parameters describing roughness ($\alpha$), brightness ($\gamma$), and number of looks ($ L $).
	The last parameter is proportional to the signal-to-noise ratio.
	The $\G$ model has attracted much attention because it characterizes areas with varying degrees of texture (from light to extreme), and the absolute value of the roughness parameter is the mean number of elementary backscatterers in the scene.
	This model is also known for its heavy-tailedness~\citep{ParameterEstimationSARStochasticDistancesKernels}, a feature that increases its flexibility and allows describing data with extreme variability.

	Several studies have proposed advances for the $\G$ distribution.
	Estimation of its parameters is often based on maximum likelihood (ML)~\citep{FreryandCribariNetoandSouza2004}, analogy~\citep{mejail2000},
	robust procedures~\citep{FrerySantAnnaMascarenhasBustos:ASP:98,mejail2000,BustosFreryLucini:Mestimators:2001,AllendeFreryetal:JSCS:05},
	bias correction~\citep{CribariFrerySilva:CSDA,VasconcellosandFreryandSilva2005,Pianto20111394}, and distance minimization~\citep{ImprovedMinimumDistanceTextureEstimatorforSpeckedData}.
	\citet{Tisonetal2004} proposed a novel {M}arkov classifier for urban areas equipped with Fisher's law reparametrized from the $\G$ distribution.
	In addition, parameter hypothesis testing based on classical inference~\citep{SilvaCribariFrery:ImprovedLikelihood:Environmetrics},
	information theory~\citep{HypothesisTestingSpeckledDataStochasticDistances},
	and non-parametric methods~\citep{ParametricNonparametricEdgeDetectionSpeckledImages,Beauchemin1998,ParametricNonparametricTestsSpeckledImagery} were proposed.
	In all these works, the observations are the only source of information.

	Regression models are tools used in data analysis and machine learning to explore and quantify the relationship between one or more measurements (predictors) and a dependent variable (the outcome).
	The primary goal of a regression model is to characterize how changes in the measurements are associated with changes in the dependent variable.
	Regression models have been used to predict biomass and soil moisture~\citep{ReviewofMachineLearningApproachesforBiomassandSoilMoistureRetrievalsfromRemoteSensingData},
	in image fusion~\citep{FusionofPolSARandPolInSARDataforLandCoverClassification},
	and in forest inventory~\citep{OntheModelAssistedRegressionEstimatorsUsingRemotelySensedAuxiliaryData}, to name a few applications.
	In addition, because of its relationship to hypothesis testing, the regression model can be used to solve change detection problems, such as using the likelihood ratio test statistic for superpixelwise  PolSAR data \citep{SuperpixelwiseLikelihoodRatioTestStatisticforPolSARDataandItsApplicationtoBuiltupAreaExtraction}.

	Quantifying the effects of latent information on $\G$-distributed SAR features is difficult.
	For instance, there are two main approaches when trying to predict intensities:
	(i)~incorporating the influence of surrounding observations,
	and (ii)~if dealing with polarimetric data, using the value from other channels.
	The former approach has been addressed by the transformation method~\citep{SARImageStatisticalModelingPartIISpatialCorrelationModelsandSimulation,GeneralizedMethodforSamplingSpatiallyCorrelatedHeterogeneousSpeckledImagery,SimulationofSpatiallyCorrelatedClutterFields} and by physical considerations~\citep{AGeneralizedGaussianCoherentScattererModelforCorrelatedSARTexture}.
	These solutions deal with spatial correlation in single-polarized data.
	The second approach relies on the multivariate nature of the polarimetric observations, for which several models have been proposed~\citep{SurveyStatisticalPolSAR}.
	A challenge with this latter approach is that the relationship between intensities is not fully understood.

	Although the density characterizing the $\mathcal G_{\text{Pol}}^0$ law is known, cf.\
	Eq.~(18) in Ref.~\citep{FreitasFreryCorreia:Environmetrics:03}, neither the joint distribution of intensities nor their correlation is known.
	Our regression model, whose response is a $\G$-distributed random variable,
	helps to understand and exploit information redundancy between intensity channels.
	This model can also be used to relate biomass and soil moisture, among other quantities of interest, to the SAR intensity return.
	As indicated by~\citet{WattsandRose2022}, our model is also suitable for describing ocean phenomena in the presence of spikes, such as Bragg waves, wind drift, and long waves.

	Few studies have attempted to use regression in SAR image features, but with different aims.
	\citet{Wang2008} proposed a K-distribution-based regression model for estimating forest biomass.
	\citet{Palm2019} proposed a Rayleigh regression model for single-look amplitude data to identify land cover and land use.
	Recently, \citet{nascimento2022} modeled K-distributed observations whose means depend on a transformation of a linear combination of regressors.
	Although the last proposal is for the same SAR feature that we use, intensity, the last model is formulated in terms of the modified Bessel function of the second kind and poses complex computational problems.
	None of these works addresses the multivariate nature of polarimetric measurements, as does our proposal.
	Our regression model, with slight adjustments in the predictor structure, is adequate for all the previous proposals.
	However, we will focus on predictors that account for continuous variables.

	There are few works in the literature about reconstructing intensity SAR from other intensities, cf.\ Refs.~\citep{Song22018,Zhao2019,Aghababaei2023}.
	Deep neural networks are used without probability assumptions about the data in these proposals.
	These representations are not interpretable in terms of multiplicative model parameters.
	Our proposal performs this reconstruction based on one of the most commonly cited and used cases of the multiplicative approach, i.e., the $\mathcal{G}^0_I$ law.
	We apply this method to reconstruct the co-polarized SAR intensity from the cross-polarized signal.
	We provide a method to verify that the generated fit is acceptable.

	In recent years, there has been increasing interest in understanding the properties of polarimetric images.
	Polarimetric SAR measurements record the amplitude and phase of backscattered signals for possible combinations of linear reception and transmission of horizontal (H) and vertical (V) polarizations.
	Since the HV and VH polarizations are usually identical,
	three different intensities are associated with channels HH, HV, and VV in each pixel of a PolSAR image.
	In this paper, we discuss how to model the intensities of SAR in one channel under the influence of the others.

	In this paper, we propose a new regression model for distributed $\G$-response variables and obtain important properties: Moments for the reciprocal $\G$ model, Fisher information matrix, two types of residuals, and four influence measures.
	In addition, we provide point and interval estimation procedures.
	We assess the performance of this model with Monte Carlo studies and actual data.
	We compare it with other regression models, namely: exponential, gamma ($\Gamma$), inverse gamma ($\Gamma^{-1}$), normal ($\mathcal{N}$), inverse normal ($\mathcal{N}^{-1}$), Weibull, power exponential, and exponential generalized beta type 2 (EGB2).
	We demonstrate the potential of our proposal in practical applications related to urban imagery scenarios.

	The paper unfolds as follows.
	Section~\ref{sec:GD} presents the main properties of the $\G$-model.
	In Section~\ref{sec:GRM}, we propose a new regression model and discuss some of its properties.
	A simulation study using the $\G$ regression parameters is presented in Section~\ref{sec:SS}.
	Applications to polarimetric data are made in Section~\ref{sec:app}.
	Section~\ref{sec:conc} discusses the main conclusions.

	\section{The $\G$ distribution}
	\label{sec:GD}

	Following the MM approach~\citep{SARImageStatisticalModelingPartISinglePixelStatisticalModels}, SAR intensities $Z\in\mathbbm{R}_+$ are defined as the product of two independent positive random variables: $Y$ describes the speckle and $X$ models the target backscatter.
	Assuming that $X \sim \Gamma^{-1}(\alpha, \gamma)$ and $Y \sim \Gamma( L , L )$, \citet{freryetal1997a} obtained the $\G$ distribution, whose probability density function is
	\begin{equation}
		\label{E:pdf}
		f_Z(z;\alpha,\gamma,L)
		=
		\displaystyle
		\frac{L^L \Gamma(L-\alpha)}{\gamma^\alpha\Gamma(-\alpha)\Gamma(L)}
		z^{L-1} (\gamma + L z)^{\alpha-L},
	\end{equation}
	where
	$-\alpha>0$,
	$\gamma>0$,
	and
	$ L >0$
	represent
	roughness,
	brightness,
	and the number of looks,
	respectively.
	This situation is denoted by $Z \sim \G(\alpha,\gamma,L)$.

	This model exhibits remarkable flexibility and interpretability.
	In particular, $|\alpha|$ measures the mean number of elementary backscatterers.
	Below, we present some results used to develop the $\G$ regression model.

	The $h$th ordinary moment of the $\G$ distribution is:
	\begin{equation}
		\E(Z^h) = \left(\displaystyle\frac{\gamma}{L}\right)^h\displaystyle\frac{B(L+h,-\alpha-h)}{B(L,-\alpha)},
		\label{MOMGI0}
	\end{equation}
	if $\alpha < -h$, where $B(\cdot,\cdot)$ is the beta function; otherwise, it diverges to infinite.

	From Equation~\eqref{MOMGI0}, one has that the variance of the $\G$-distributed random variable is
	\begin{equation}
		\label{varZgi0}
		\text{Var}(Z)=\mu^2\Big[\Big(\displaystyle\frac{\alpha+1}{\alpha+2}\Big)\displaystyle\frac{L+1}{L}-1\Big],
	\end{equation}
	where $\mu= \E(Z)$, if $\alpha < -2$; otherwise it diverges to infinite.
	Hereafter, we assume $\alpha < -2$.
	Note that an increase in $\alpha$ or $\mu$ at SAR indicates scenes with higher variability, while an increase in $L$ implies areas with better signal-to-noise ratios.
	The following two results are contributions of this work.

	\begin{COR}\label{C:EV_T}
		Let $Z \sim \G(\alpha,\gamma,L)$.
		The $h$th reciprocal moment of $T=\gamma+L Z$ is given by
		$$
		\E\Big[\displaystyle\frac{1}{T^h}\Big] = \displaystyle\frac{1}{\gamma^h} \displaystyle\frac{B(-\alpha+h,L)}{B(-\alpha,L)}.
		$$
	\end{COR}

	\begin{LEM}\label{L:SF}
		The $\G$ distribution is a scale family.
	\end{LEM}
	Appendices~\ref{App2} and~\ref{App3} provide the proofs.

	\section{$\G$ Regression Model}
	\label{sec:GRM}

	The discussion in this section focuses on using the regression model for inferring the intensities in a polarimetric channel, given the data in other channels.
	The reader must remember that the proposed regression model can be used to describe the observed intensity as a function of any linear combination of explanatory variables.

	The multivariate nature of polarimetric information involves the notion of redundancy: each component (or channel) retrieves a portion of the information in the entire signal.
	For example, features from the HH channel may have a different correlation structure with the HV and VV channels.
	In this section, we present a mathematical treatment to study the linear dependence between the intensities of the polarization channels.

	From Lemma~\ref{L:SF} one can define a linear model for $\G$-distributed random variables.
	Let $Z_k$ be such a random variable:
	\begin{align}
		Z_k = \epsilon_k \exp({{\bm x}_k^\top{\bm \beta}}) ,
		\text{ for }
		\quad
		k=1,\ldots,n,
		\label{E:GLM}
	\end{align}
	where
	${\bm \beta}=(\beta_0,  \beta_1, $ $ \ldots,  \beta_p)^\top$ is a vector of regression coefficients,
	${\bm x}_k =$ $ (1,  x_{k 1},  \ldots,  x_{k p})^\top$
	is the vector of $p < n$ explanatory variables (assumed fixed and known),
	and
	$\epsilon_k \sim \G(\alpha,  -(\alpha+1),   L )$.
	Thus, it follows from Lemma~\ref{L:SF} that
	$$Z_k\mid {\bm x}_k\sim  \G(\alpha, \gamma_k,  L ),
	$$
	where
	$\gamma_k = \mu_k (-\alpha-1)$,
	$\mu_k = g^{-1}(\eta_k)$,
	$\eta_k={\bm x}_k^\top{\bm \beta}$,
	and
	$g(\cdot)$ is a positive, strictly monotone, and twofold differentiable link function.
	There are several traceable link functions; for example~\citep{McCullagh1989}:
	\begin{itemize}
		\item Extended logit: $g(\mu_k) = \log \frac{F_{(0,\infty)}(\mu_k)}{1-F_{(0,\infty)}(\mu_k)}$,
		\item Complementary log-log: $g(\mu_k)=\log(-\log(1-F_{(0,\infty)}(\mu_k)))$, and
		\item Logarithmic: $g(\mu_k)=\log\mu_k$,
	\end{itemize}
	where $F_{(0,\infty)}(\cdot)$ is the cumulative distribution function of any distribution with positive support.
	We choose $g(\mu_k)=\log\mu_k$ because of its analytical tractability and because it transforms the multiplicative speckle effect into an additive one.

	Thus, we assume that $n$ SAR intensities in a region from a given channel are described by random variables $Z_1, \ldots, Z_n$ such that $Z_k \sim \G(\alpha, \gamma_k, L)$
	with mean $\mu_k=\exp({{\bm x}_k^\top{\bm \beta}})$, where ${\bm x}_k$ is the known vector of features that can be associated with $Z_k$; e.g., a biomass index \citep{Wang2008}, other features on the same channel (such as phase), and the same features from the other channels.
	In this work, we addressed the letter case: mapping the linear dependence/influence of one channel on the others.

\begin{figure}[!htbp]
		\centering
		\subfigure[San Francisco image \label{ilu1}]{\includegraphics[width=.48\linewidth]{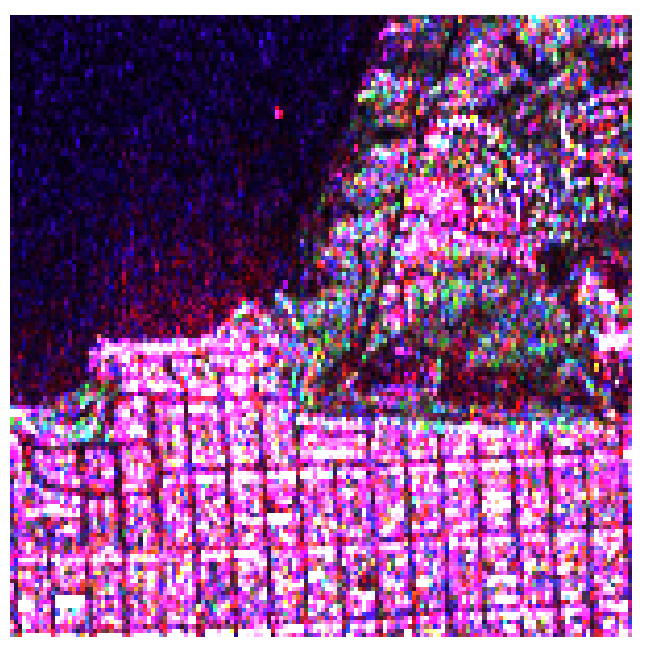}}
		\subfigure[Map of $\widehat\alpha$\label{ilu3}]{\includegraphics[width=.48\linewidth]{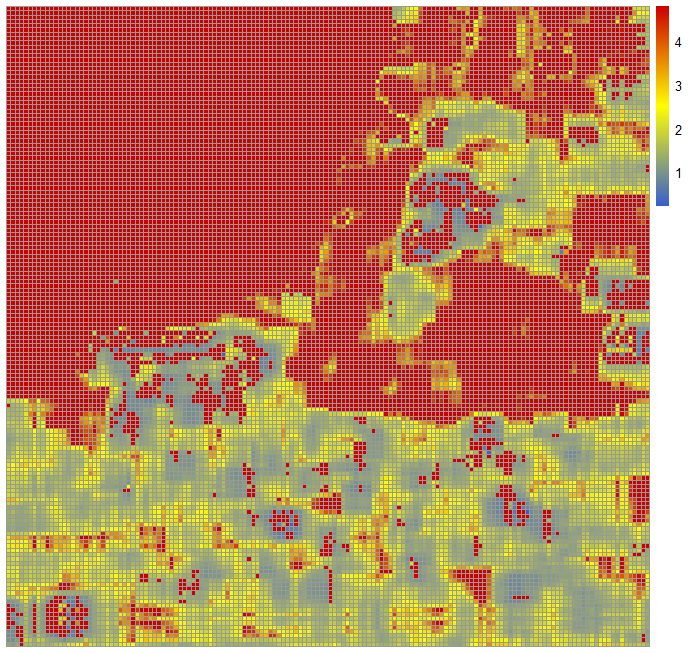}}
		\\
		\subfigure[Map of $\widehat\gamma$\label{ilu2}]{\includegraphics[width=.48\linewidth]{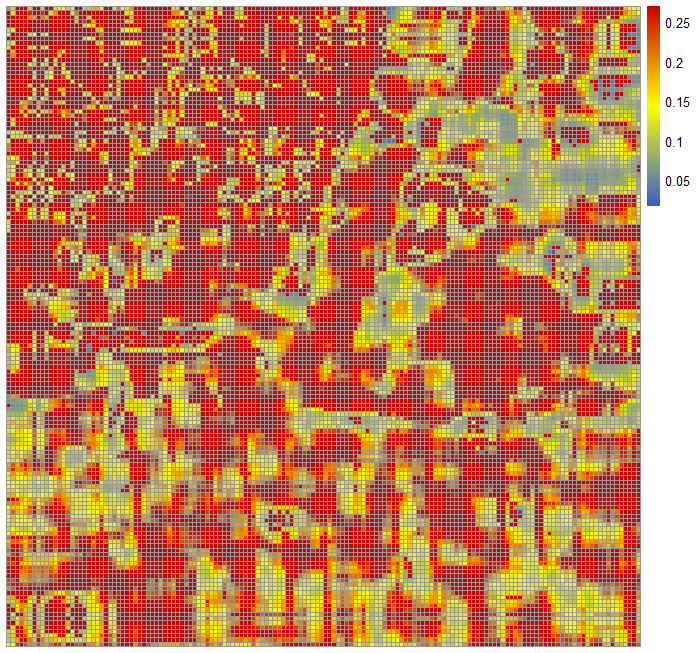}}
		\subfigure[Map of $\widehat\mu$\label{ilu4}]{\includegraphics[width=.48\linewidth]{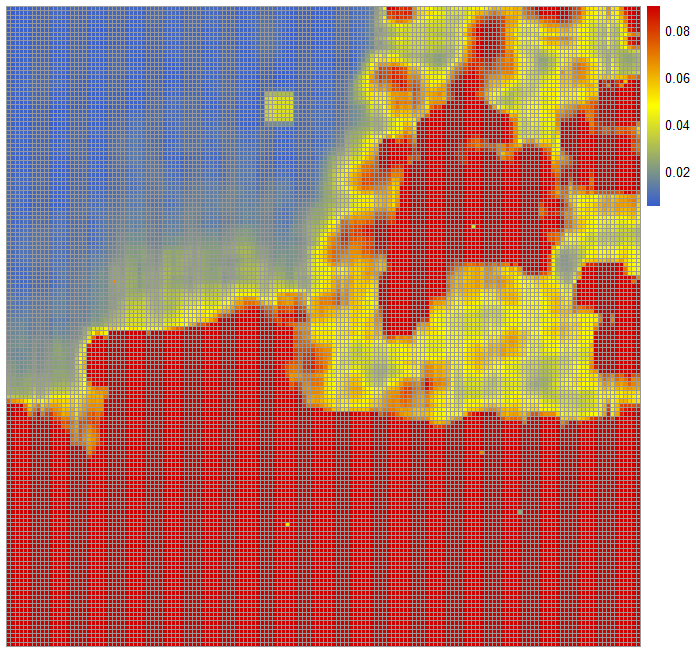}}
		\caption{Maps of local estimates $\alpha$, $\gamma$, and $\mu$.}
		\label{Apli1}
	\end{figure}

	To illustrate the use of the regression model, consider the AIRSAR image over San Francisco (USA) with four nominal looks and resolution \qtyproduct{10x10}{\meter} shown in Fig.~\ref{ilu1}.
	The HH, HV, and VV intensities were mapped to the R, G, and B channels, respectively, to produce a false-color image.
	This scene consists of three outstanding regions: Sea, Forest, and City.
	To assess the effect of region type on the maximum likelihood estimates of $\alpha,\gamma,$ and $\mu$, we present maps of these estimates obtained on $7 \times 7$ windows for each pixel.
	The $\widehat{\alpha}$ map (Fig.~\ref{ilu3}) refers to local roughness and highlights urban patches.
	The $\widehat{\gamma}$ map (Fig.~\ref{ilu2}) measures the local scale that, along with the values of $\widehat{\alpha}$, gives the map of local mean values (Fig.~\ref{ilu4}).

	\subsection{Estimation by maximum likelihood}

	Let
	$(Z_1\mid\bm{x}_1), \ldots, (Z_n\mid\bm{x}_n)$
	be an $n$-point random sample such that $Z_k\sim\G(\alpha,\gamma_k,L)$
	and $\gamma_k= (-\alpha-1)\exp({\bm{x}_k^\top\bm{\beta}})$.
	From~\eqref{E:pdf} and~\eqref{E:GLM},
	the log-likelihood function at
	${\bm \theta}=({\bm \beta}^\top, \alpha,L)^\top$
	is given as
	\begin{equation}\label{E:LL}
		\ell({\bm \theta})=\displaystyle\sum_{k=1}^{n}\ell_k({\bm\theta}),
	\end{equation}
	where
	\begin{align*}
		\ell_k({\bm \theta}) =
		L \log L+\log\Gamma(L-\alpha)-\alpha\log\mu_k -\alpha\log(-\alpha-1)
		- \log\Gamma(-\alpha) - \log\Gamma(L)
		+(L-1)\log z_k+(\alpha-L)\log (\gamma_k +  L  z_k),
	\end{align*}
	such that
	$z_k$ is an outcome from $Z_k$.

	Based on~\eqref{E:LL}, the associated score function,
	\begin{equation*}
		U({\bm \theta})=(U_{\bm \beta},U_\alpha,U_ L)
		=
		\Big(
		\displaystyle\frac{\partial \ell({\bm \theta})}{\partial{\bm \beta}^\top},
		\displaystyle\frac{\partial \ell({\bm \theta})}{\partial\alpha},
		\displaystyle\frac{\partial \ell({\bm \theta})}{\partial L }
		\Big)^\top,
	\end{equation*}
	is determined by:
	\begin{itemize}

		\item $U_{\bm \beta}$:
		\begin{equation}\label{E:MLEb}
			U_{\bm \beta} = \alpha{\bm X}^{\top}\bm{E}({\bm T}^\ast - {\bm \mu}^\ast),
		\end{equation}
		where ${\bm X}$ is an $n \times p$ matrix whose $k$th row is ${\bm x}_k^{\top}$,
		$$
		\bm{E} = \text{diag}
		\{
		{1}/{g^\prime(\mu_1)},
		\ldots,
		{1}/{g^\prime(\mu_k)},
		\ldots,
		{1}/{g^\prime(\mu_n)}
		\},
		$$
		$g^\prime(x)=\operatorname{d} g(x)/\operatorname{d}x$,
		${\bm T}^\ast = (T_1^\ast, \ldots, T_k^\ast, \ldots, T_n^\ast)^\top$,
		with
		$T_k^\ast = \sfrac{-c_1}{T_k}$,
		$c_1 = ( L  - \alpha)(-\alpha-1),$
		and
		${\bm \mu}^\ast = ({1}/{\mu_1}, \dots,
		{1}/{\mu_k}, \dots, {1}/{\mu_n})^\top$;
		\item[]
		\item $U_\alpha$:
		\begin{equation}
			\label{E:MLEa}
			U_{\alpha} =
			n U_1(\alpha,L) + \displaystyle\sum_{k=1}^{n}\log \displaystyle\frac{T_k}{\mu_k}
			-(\alpha-L)\displaystyle\sum_{k=1}^{n}\displaystyle\frac{\mu_k}{T_k},
		\end{equation}
		where
		\begin{align*}
			U_1(\alpha,L) =-\Psi(L-\alpha)+\Psi(-\alpha)
			- \log(-\alpha-1) + \displaystyle\frac{\alpha}{(-\alpha-1)},
		\end{align*}
		and
		$\Psi(x)=\operatorname{d}\log \Gamma(x)/\operatorname{d}x$ is the digamma function;

		\item[]
		\item $U_L$:
		\begin{equation}
			\label{E:MLEl}
			U_L= n U_2(\alpha, L) +
			\displaystyle\sum_{k=1}^{n}\log\displaystyle\frac{z_k}{T_k}
			+ (\alpha-L)\displaystyle\sum_{k=1}^{n}\displaystyle\frac{z_k}{T_k},
		\end{equation}
		where
		$$
		U_2(\alpha,L)=1+\log L+\Psi(L-\alpha)-\Psi(L).
		$$

	\end{itemize}

	In this case,
	the Fisher information matrix is given by
	\begin{equation}\label{E:FIM}
		{\bm K}({\bm\theta})
		=
		\E\Bigl[
		-
		\frac{
			\partial^2 \ell({\bm \theta})
		}{
			\partial {\bm \theta}^\top \partial {\bm \theta}
		}
		\Bigr]
		=
		\begin{bmatrix}
			{\bm K}_{{\bm \beta} {\bm \beta}} & {\bm K}_{{\bm \beta} \alpha}  & {\bm K}_{{\bm \beta}  L }
			\\[3mm]
			{\bm K}_{{\bm \beta} \alpha}     & {\bm K}_{\alpha \alpha}       & {\bm K}_{\alpha  L }
			\\[3mm]
			{\bm K}_{{\bm \beta}  L }                           & {\bm K}_{\alpha  L }                      & {\bm K}_{ L   L }
			\\
		\end{bmatrix},
	\end{equation}
	where
	${\bm K}_{{\bm \beta} {\bm \beta}}= \alpha{\bm X}^\top{\bm W}{\bm X}$,
	${\bm K}_{{\bm \beta} \alpha} = c_2{\bm X}^\top \bm{E}{\bm \mu}^\ast$,
	${\bm K}_{\alpha \alpha} = nc_3$,
	${\bm K}_{ L   L } = nc_4$,
	${\bm K}_{{\bm \beta}  L } = {\bm K}_{\alpha  L } = {\bm 0}$,
	${\bm W} = \text{diag}
	\left\{\omega_1, \ldots, \omega_t, \ldots, \omega_n\right\}$,
	and
	$$
	\omega_k = \left(\displaystyle\frac{ L }{ L -\alpha+1}\right)
	\displaystyle\frac{1}{\mu_k^2 [g^\prime(\mu_k)]^2}.
	$$
	For a detailed discussion of the elements of ${\bm K}({\bm \theta})$ and $c_2, c_3$, and $c_4$, see Appendix~\ref{App4}.
	Note that the parameters ${\bm \beta}$ and $\alpha$ are not orthogonal (in the sense that $\theta_i$ and $\theta_j$ are orthogonal in $\bm{\theta}=(\theta_1,\ldots,\theta_p)^\top$, denoted by ``$\theta_i\perp \theta_j$'', if only $\E[\partial^2 \ell(\bm{\theta})/\partial \theta_i\partial \theta_j]=0$ for all $i\neq j$~\citep{huzurbazar1950}), which is the opposite of what is proved in the class of generalized linear regression models \citep[see ][]{McCullagh1989}.
	On the other hand, ${\bm \beta} \perp L $ and $\alpha \perp L $.

	For a sufficiently large sample and under the usual regularity conditions~\citep{Bickel2001}, it follows that
	\begin{align*}
		\sqrt{n} ( \widehat{\bm \theta}  -  {\bm \theta}  )
		\overset{\mathcal{D}}{\underset{n \to \infty}{\longrightarrow}}
		\No_{p+1}\left({\bm 0}_{p+1}, {\bm K}^{-1}({\bm \theta})\right),
	\end{align*}
	where ${\bm 0}_{k}$ is the $k$-dimensional column vector of zeros,
	$\widehat{\bm \theta}$ is the maximum likelihood estimator (MLE) for ${\bm \theta}$,
	$\overset{\mathcal{D}}{\underset{n \to \infty}{\longrightarrow}}$ denotes convergence in distribution,
	and $\No_{p}({\bm \mu}, {\bm \Sigma})$ indicates the multivariate normal distribution with mean ${\bm \mu}$ and covariance matrix ${\bm \Sigma}$.
	In this paper, we obtain a closed-form expression for ${\bm K}^{-1}({\bm \theta})$ of the $\G$ regression model:
	\begin{equation}\label{E:IFIM}
		{\bm K}^{-1}({\bm \theta}) =
		\begin{bmatrix}
			{\bm K}^{{\bm \beta} {\bm \beta}} & {\bm K}^{{\bm \beta} \alpha}  & {\bm K}^{{\bm \beta}  L }
			\\
			{\bm K}^{{\bm \beta} \alpha}   & {\bm K}^{\alpha \alpha}       & {\bm K}^{\alpha  L }
			\\
			{\bm K}^{{\bm \beta}  L }                           & {\bm K}^{\alpha  L }                         & {\bm K}^{ L   L }
		\end{bmatrix},
	\end{equation}
	with elements given in Appendix~\ref{App4}.

	The asymptotic multivariate normal distribution can also be used to derive asymptotic confidence intervals, which are useful for testing the significance of submodels and comparing them using likelihood ratio, score, and Wald statistics.
	In particular, the asymptotic confidence interval at $100 (1-\epsilon)\%$ for the $k$th component of ${\bm \theta}$, $\theta_k$, is given by
	\begin{align*}
		\mbox{ACI}_{\epsilon}\left(\theta_k\right)
		&= \left(\widehat{\theta}_k - z_{\epsilon/2}\sqrt{\widehat{{\bm K}}^{k, k}} ;
		\widehat{\theta}_k + z_{\epsilon/2}\sqrt{\widehat{{\bm K}}^{k, k}}\right),
	\end{align*}
	where $\widehat{{\bm K}}^{k,k}$ denotes the $k$th element of the main diagonal of ${\bm K}^{-1}({\bm \theta})$ and $z_{\epsilon/2}$ is the $(1-\epsilon/2)$ quantile of the standard normal distribution.

	We obtain the MLE for $\alpha$, $ L $, and ${\bm \beta}$ by maximizing the log-likelihood function by numerically solving the system $(U_{\bm \beta}, U_\alpha, U_ L )=(0,0,0)$.
	We used iterative methods such as BFGS (Broyden-Fletcher-Goldfarb-Shanno) or Newton-Raphson~\citep{Nocedal1999}.
	These optimization algorithms require initial estimates, and we suggest using ordinary least squares estimates obtained from a linear regression model of the transformed responses
	$g(z_1), g(z_2), \ldots, g(z_n)$;
	i.e.,
	$({\bm X}^\top {\bm X})^{-1}{\bm X}^\top \log({\bm z})$,
	where $\log({\bm z})=[$ $\log(z_1),$ $\ldots,$ $\log(z_k),$ $\ldots,$ $\log(z_n)]^\top$.
	For the preliminary estimation of $\alpha$, the value $ L $ was determined, which equates the score components of~\eqref{E:pdf} with the null vector, from which follows
	\begin{align}
		\label{E:derA}
		U_{\alpha}&=n[\Psi(-\alpha)-\Psi(L-\alpha) - \log(\gamma)]+\sum_{k=1}^n\log T_k=0,
		\intertext{and that}
		\label{E:derG}
		U_{\gamma}&=-\frac{n\alpha}{\gamma}+(\alpha- L)\sum_{k=1}^n\frac{1}{T_k}=0.
	\end{align}
	From \eqref{E:derG},
	we have the initial guess of $\alpha$ given by
	$\alpha=L\frac{S}{S-n/\gamma}$,
	where
	$S=\sum_{k=1}^n T_k^{-1}$,
	and $\gamma$
	can be obtained by solving the nonlinear equation
	\begin{equation*}
		\frac{1}{n}\sum_{k=1}^{n}\log\frac{T_K}{\gamma}=\Psi\Big(\frac{L S}{S-n/\gamma}\Big)-\Psi\Big(L\Big(1+\frac{S}{S-n/\gamma}\Big)\Big).
	\end{equation*}

	\subsection{
		Residual analysis
	}

	An essential step in proposing a new regression model is to analyze its assumptions using the residuals.
	For this purpose, we introduce methods for two classes of residuals.

	Residual analyses are often conducted in the form of ordinary residuals, standardized variants, or variance residuals.
	They serve a dual purpose: to identify outliers (the sensitivity or influence study) and test the model assumptions.
	Note that the first function can be used to identify atypical phenomena in image processing:
	such as corner reflectors in urban areas~\citep{GrootandOtten1994} and the effect of Bragg waves in marine areas~\citep{WattsandRose2022}.

	In this case, it is the $k$th standardized residual:
	$$
	r_k = \displaystyle\frac{Z_k-\widehat{\mu}_k}{\sqrt{\widehat{\text{Var}}(Z_k)}},
	$$
	where
	$\widehat{\mu}_k = g^{-1}(x^\top_k\widehat{\bm \beta})$,
	and the estimate for variance of $Z_k$ is
	$$
	\widehat{\operatorname{Var}}(Z_k) =
	\widehat{\mu}_k^2\Big[
	\Big(\frac{\widehat{\alpha}+1}{\widehat{\alpha}+2}
	\Big)\frac{L+1}{L}-1\Big].
	$$

	A second residual is the deviance residual~\citep{HardinHibe2012, Amin2016}.
	The term deviance refers to a tool for quantifying the difference between a full/saturated model and a reduced/used model, using the likelihood as the criterion.
	By the following definition, a point with a large residual makes the obtained coefficients ``look bad,'' i.e., it explains much of the discrepancy between our model and the saturated model.
	Applying this to the case of $\G$-regression, we get:
	$$\text{\bf D}_{d k}({\bm z};
	{\bm \mu}, \alpha,L) = \operatorname{sign}(z_k-\mu_k)\sqrt{\Big|2
		\Big[
		\alpha\log\Big(\frac{z_k}{\mu_k}T_k^\diamond \Big) -
		L  \log\left(T_k^\diamond\right)
		\Big]\Big|}$$
	as a residual,
	where
	$T_k^\diamond = \sfrac{T_k}{z_k(-\alpha-1 +  L )}$
	and
	$\operatorname{sign}(\cdot)$
	is the signum function defined as
	\[
	\operatorname{sign}(x)
	=
	\begin{cases}
		+1, & \text{if $x>0$,}
		\\
		0, & \text{if $x=0$,}
		\\
		-1, & \text{if $x<0$.}
	\end{cases}
	\]
	Hence, the standardized deviance residuals are defined by~\citet{Davison1989} as
	\[SR_{k}=\frac{\text{\bf D}_{d k}({\bm z};{\bm \mu},\alpha,L)}
	{S\sqrt{1-h_{k k}}},
	\]
	where
	$S = \sum_{k=1}^{n}
	\sfrac{\text{\bf D}_{d k}({\bm z};{\bm \mu},\alpha,L)^2}
	{(n - p)}$.

	When the distribution of residuals is not known, normal plots with simulated envelopes are a useful diagnostic tool, cf.~\citet[section~4.2]{Atkinson1985} and \citet[section~14.6]{Neter1996}.
	Simulated envelopes can be used to decide whether the observed residuals are consistent with the fitted model.

	We need the minimum and maximum values of the $\nu$ order statistics to create the envelope.
	We follow \citet[p.~36]{Atkinson1985} who suggested $\nu=19$, and the probability that a given absolute residual falls above the upper band provided by the envelope is approximately equal to $0.05$.
	As a decision rule, if some absolute residuals fall outside the bounds provided by the simulated envelope, we must examine these residuals.
	On the other hand, if many points lie outside this range, this argues against the suitability of the fitted model.

	\subsection{Influential measures}

	After discussing the types of residuals, it is necessary to define measures for identifying influential observations, which is the topic of this section.
	In what follows, we derive four influential measures: a generalized leverage measure, a projection-based measure (also called a hat matrix), Cook distance, and difference in fits (DFFITS).

	\citet{Wei1998} developed the generalized leverage, which is defined as
	\begin{equation}\label{E:GL1}
		\text{\bf GL}(\widetilde{\bm \theta}) =
		\displaystyle\frac{\partial \widetilde{\bm z}}{\partial {\bm z}^\top},
	\end{equation}
	where ${\bm \theta}$ is a parameter vector such that $\E({\bm z})=\bm{\mu}({\bm \theta})$, $\widetilde{\bm \theta}$ is an estimator of ${\bm \theta}$, and $\widetilde{\bm z} = \mu(\widetilde{\bm \theta})$.
	Let $\ell({\bm \theta})$ be the $\G$ log-likelihood function with continuous second order derivatives with respect to ${\bm \theta}$ and ${\bm z}$.
	Also, let $\widehat{\bm \theta}$ be the MLE for ${\bm \theta}$, assuming it exists and is unique, and let $\widehat{\bm z}$ be the predicted response vector.
	\citet{Wei1998} showed that the generalized leverage $n \times n$ matrix in~\eqref{E:GL1} can be expressed by
	$$
	\text{\bf GL}({\bm \theta}) =
	D_{\bm \theta}
	\Big[-\displaystyle\frac{\partial^2 \ell({\bm \theta})}
	{\partial {\bm \theta} \partial {\bm \theta}^\top}\Big]^{-1}
	D_{\bm \theta  z},
	$$
	evaluated at $\widehat{\bm \theta}$, where $D_{\bm \theta} = \displaystyle\frac{\partial \E({\bm z})}{\partial {\bm \theta}^\top}$ and $D_{\bm \theta  z} = \displaystyle\frac{\partial^2 \ell({\bm \theta})}{\partial {\bm \theta} \partial {\bm z}^\top}.$

	Considering $\alpha$ as a nuisance parameter and $L$ fixed, we obtained a closed-form expression for {\bf GL }$({\bm \beta})$ in the $\G$ regression model.
	We have:
	\begin{itemize}
		\item[(i)]
		$
		D_{\bm \beta} = \bm{E}{\bm X},
		$
		\item[(ii)]
		$
		-\displaystyle\frac{\partial^2 \ell({\bm \theta})}
		{\partial {\bm \beta} \partial {\bm \beta}^\top} =
		\alpha{\bm X}^\top{\bm Q}{\bm X},
		$
		where
		${\bm Q}= \text{diag}\left\{ q_1, \ldots, q_k, \ldots, q_n \right\}$
		with
		$$
		q_t = -
		\frac{
			\displaystyle\frac{1}{\mu_k^2} + \frac{c_1(-\alpha-1)}{\alpha T_k^2}
			+ \Big[
			\frac{1}{\mu_k} + \frac{c_1}{\alpha T_k}
			\Big]\frac{g^{\prime \prime}(\mu_k)}{g^{\prime}(\mu_k)}}
		{[g^{\prime}(\mu_k)]^2},
		$$
		and
		\item[(iii)]
		$
		\displaystyle\frac{\partial^2 \ell({\bm \theta})}
		{\partial {\bm \beta} \partial {\bm z}^\top} =
		\alpha{\bm X}^\top\bm{E}{\bm T}^\ast,
		$
		where
		$
		{\bm T}^\ast
		=
		\frac{ L  c_1}{\alpha}
		\text{diag}\left\{
		1/{T_1^2}, \ldots,
		1/{T_k^2}, \ldots,
		1/{T_n^2} \right\}
		$.
	\end{itemize}
	Therefore, the following identity holds:
	\begin{equation}\label{E:GL}
		\text{\bf GL}({\bm \beta})
		=
		\bm{E}{\bm X}({\bm X}^\top{\bm Q}{\bm X})^{-1}{\bm X}^\top\bm{E}{\bm T}^\ast.
	\end{equation}
	More details are presented in Appendix~\ref{App5}.

	Replacing the observed information matrix with the Fisher information matrix, Expression~\eqref{E:GL} collapses to
	$$
	\text{\bf GL}_{\text{\tiny A}}({\bm \beta}) =
	\bm{E}{\bm X}({\bm X}^\top{\bm W}{\bm X})^{-1}{\bm X}^\top\bm{E}{\bm T}^\ast.
	$$
	Hence, the $(k, k)$-entry of
	$\text{\bf GL}_{k k}$
	can be expressed as
	$$
	\text{\bf GL}_{k k}({\bm \beta}) =
	\omega_k x_k^\top({\bm X}^\top{\bm W}{\bm X})^{-1}x_k,
	$$
	where
	\begin{equation}\label{E:wt}
		\omega_k = \frac{\partial^2 \ell_k({\bm \theta})}{\partial \mu_k^2}
		\left(\frac{\partial \mu_k}{\partial T_k}\right)^2
		= \left[\frac{\alpha}{\mu_k^2} +
		\frac{c_1(-\alpha-1)}{T_k^2}\right]
		\frac{1}{{g^\prime(\mu_k)}^2}.
	\end{equation}

	\citet{Pregibon1981} extended the projection/hat matrix from linear regression $\widehat{z}$ against $X$ under weights to generalized linear models:
	$$
	\text{\bf H} =
	{\bm W}^{\sfrac{1}{2}}{\bm X}({\bm X}^\top{\bm W}{\bm X})^{-1}
	{\bm X}^\top{\bm W}^{{1}/{2}}.
	$$
	In the literature, the use of the $(k, k)$-entry of $\text{\bf H}$, say $h_{k k}$, has been proposed to detect the presence of leverage points in linear generalized regression models~\citep{Wei1998,Lindsey1997}.
	It is worth noting that $\text{\bf GL}$ and $\text{\bf H}$ coincide for large samples.

	Assume $\alpha$ and $ L $ are unknown,
	hence ${\bm \theta}^\top = ({\bm \beta}^\top, \alpha,  L )$.
	Let $D_{\theta}=[({\bf E}{\bf X})^\top\quad {\bm 0}_n^\top\quad {\bm 0}_n^\top]^\top$.
	Moreover,
	\begin{equation*}
		\frac{\partial^2 \ell({\bm \theta})}
		{\partial {\bm \theta} \partial {\bm z}^\top} =
		\begin{bmatrix}
			\alpha{\bm X}^\top\bm{E}{\bm T^\ast}
			\\
			{\bm a}^\top
			\\
			{\bm b}^\top
		\end{bmatrix},
	\end{equation*}
	where ${\bm a}=(a_1, \ldots, a_k, \ldots, a_n)^\top$ with
	$a_k={L[T_k+(\alpha-L)\mu_k]}/{T_k^2}$, and
	${\bm b}=(b_1, \ldots, b_k, \ldots, b_n)^\top$ with
	$b_k={1}/{z_k}-{L}/{T_k}+
	(1-{L  z_k}/{T_k}){(\alpha-L)}/{T_k} .$
	Finally, using on $-{\partial^2 \ell({\bm \theta})}/{\partial {\bm \theta} \partial {\bm \theta}^\top}$ the same idea on which Expression~\eqref{E:FIM} was built, we have that
	\begin{equation}
		\label{E:der}
		-\displaystyle\frac{\partial^2 \ell({\bm \theta})}
		{\partial {\bm \theta} \partial {\bm \theta}^\top} =
		\begin{bmatrix}
			\alpha{\bm X}^\top{\bm Q}{\bm X} & {\bm X}^\top \bm{E}{\bm M} & {\bm X}^\top \bm{E}{\bm N}
			\\
			{\bm X}^\top \bm{E}{\bm M}                         & {\bm R}                     & {\bm P}
			\\
			{\bm X}^\top \bm{E}{\bm N}                          & {\bm P}                    & {\bm S} \\
		\end{bmatrix},
	\end{equation}
	where
	${\bm M} = \text{diag}\{m_1, \ldots, m_k, \ldots, m_n\}$
	with
	$$m_k = \displaystyle\frac{1}{\mu_k} +
	\frac{1}{T_k}\Big(2\alpha- L +1 + \frac{\mu_k c_1}{T_k}\Big),$$
	${\bm N} = \text{diag}\{n_1, \ldots, n_k, \ldots, n_n\}$
	with
	$$n_{k}=\displaystyle\frac{(-\alpha-1)}{T_k}
	\Big[1 + \frac{(\alpha -  L )z_k}{T_k}\Big],$$
	${\bm P} = \operatorname{tr}({\bm P}^\ast),$
	where ${\bm P}^\ast=
	\text{diag}\{p_1^\ast, \ldots, p_k^\ast, \ldots, p_n^\ast\}$
	such that $\operatorname{tr}(\cdot)$ represents the trace of a matrix argument
	with
	$$
	p_k^\ast=\Psi^{(1)}(L-\alpha)+\frac{1}{T_k}(z_k+\mu_k)+\frac{(\alpha-L)\mu_k z_k}{T_k^2},
	$$
	${\bm R}=\operatorname{tr}({\bm R}^\ast),$
	where
	${\bm R}^\ast=\text{diag}\{r_1^\ast, \ldots, r_k^\ast, \ldots, r_n^\ast\}$
	with
	$$
	r_k^\ast=-U_1^{(1)}(\alpha,L)+\frac{\mu_k}{T_k}\Big[2+(\alpha-L)\frac{\mu_k}{T_k}\Big],
	$$
	and
	${\bm S} = \operatorname{tr}({\bm S}^\ast),$
	where
	${\bm S}^\ast = \text{diag}\{s_1^\ast, \ldots, s_k^\ast, \ldots, s_n^\ast\}$
	with
	$$
	s_k^\ast
	=
	-\Psi^{(1)}(L-\alpha)+\Psi^{(1)}(L)-\frac{1}{L}+\frac{z_k}{T_k}\Big[2+(\alpha-L)\frac{z_k}{T_k}\Big],
	$$
	where
	$\Psi^{(k)}(x)={\partial^{k+1}\log \Gamma(x)}/{\partial x^{k+1}}$ for $x > 0.$
	Now, partitioning matrix~\eqref{E:der}, we have that the inverse of  ${\bm K}({\bm \theta})$ is given by
	\begin{equation*}
		\left(-\displaystyle\frac{\partial^2 \ell({\bm \theta})}
		{\partial {\bm \theta} \partial {\bm \theta}^\top}\right)^{-1} =
		\begin{bmatrix}
			{\bm A}^\ast & {\bm B}^\ast & {\bm C}^\ast
			\\
			{\bm B}^\ast      & {\bm D}^\ast & \bm{E}^\ast
			\\
			{\bm C}^\ast      & \bm{E}^\ast      & {\bm F}^\ast
		\end{bmatrix},
	\end{equation*}
	where
	${\bm A}^\ast = \left[(\alpha{\bm X}^\top{\bm Q}{\bm X})^{-1}
	+ {\bm \zeta}^\ast {\bm \vartheta}^{\ast^{-1}} {\bm \zeta}^{\ast^\top}\right]
	{\bm \Upsilon}^\ast {\bm  \Phi}^{\ast^{-1}} {\bm \Upsilon}^{\ast^\top}$,
	${\bm B}^\ast = \left[-{\bm \zeta}^\ast {\bm \vartheta}^{\ast^{-1}}\right]$ $
	{\bm \Upsilon}^\ast {\bm  \Phi}^{\ast^{-1}} {\bm \Upsilon}^{\ast^\top}$,
	${\bm D}^\ast = {\bm \vartheta}^{\ast^{-1}}
	{\bm \Upsilon}^\ast {\bm  \Phi}^{\ast^{-1}} {\bm \Upsilon}^{\ast^\top}$,
	${\bm C}^\ast =
	{\bm \Upsilon}^\ast {\bm  \Phi}_1^{\ast^{-1}} {\bm \Upsilon}^{\ast^\top}$,
	$\bm{E}^\ast =
	{\bm \Upsilon}^\ast {\bm  \Phi}_2^{\ast^{-1}} {\bm \Upsilon}^{\ast^\top}$,
	${\bm F}^\ast = {\bm  \Phi}^{\ast^{-1}},$
	${\bm \vartheta} = {\bm R} -  ({\bm X}^\top \bm{E}{\bm N})^\top
	(\alpha{\bm X}^\top{\bm Q}{\bm X})^{-1}({\bm X}^\top \bm{E}{\bm N})$,
	${\bm \zeta}^\ast =
	(\alpha{\bm X}^\top{\bm Q}{\bm X})^{-1}$ $({\bm X}^\top \bm{E}{\bm N})$,
	${\bm \Upsilon}^\ast = {\bm A}^{\star^{-1}}{\bm B}^{\star}$,
	${\bm  \Phi}^{\ast} = {\bm D}^{\star} - {\bm B}^{\star^\top}{\bm \Upsilon}^\ast$,
	${\bm  \Phi}^{\ast} = [  {\bm  \Phi}_1^{\ast} \quad {\bm  \Phi}_2^{\ast} ]$.
	Therefore, it can be shown that
	$$
	\text{\bf GL}({\bm \beta}, \alpha,  L ) = \text{\bf GL}({\bm \beta}) +
	\bm{E}{\bm X}{\bm B}^{\ast}{\bm a}^\top +
	\bm{E}{\bm X}{\bm C}^{\ast}{\bm b}^\top,
	$$
	where $\text{\bf GL}({\bm \beta})$ is given in~\eqref{E:GL}.
	Notice that, for sufficiently large values of $|\alpha|$, $\text{\bf GL}({\bm \beta}, \alpha,  L ) \rightarrow
	\text{\bf GL}({\bm \beta})$ holds true.

	\citet{Cook1977} developed a measure of influence for each observation.
	This distance quantifies the effect of the $t$th observation by the squared distance between $\widehat{\bm \beta}$ and $\widehat{\bm \beta}_{(t)}$ (where $\widehat{\bm \beta}_{(t)}$
	is the parameter estimate without the $t$th observation).
	The Cook distance is given by
	$$
	\text{\bf D}_c({\bm \theta})
	=
	{\frac 1p}
	(\widehat{\bm \beta} - \widehat{\bm \beta}_{(k)})^\top
	{\bm X}{\bm W}{\bm X}
	(\widehat{\bm \beta} - \widehat{\bm \beta}_{(k)}).
	$$
	We can use its usual approximation:
	$$
	\text{\bf D}_c^\ast({\bm \theta})
	=
	\frac{h_{k k}r_k^2}{p(1-h_{k k})^2},
	$$
	which spares us the $n+1$ adjustment of the model and combines leverage and residuals \citep{Cook1986}.

	We also determined the DFFITS diagnostic measure proposed by \citet{White1980} that shows how influential a point is in linear regression.
	This measure is defined as
	$$
	\text{\bf DFFITS}({\bm \theta}) =
	\big(\widehat{\bm \beta}-\widehat{\bm \beta}_{(k)}\big)
	\sqrt{\frac{h_{k k}}{1-h_{k k}}}.
	$$

\begin{figure}[!htbp]
		\centering
		\subfigure[\label{F:estSim} A fitted line plot]{\includegraphics[width=.4\linewidth]{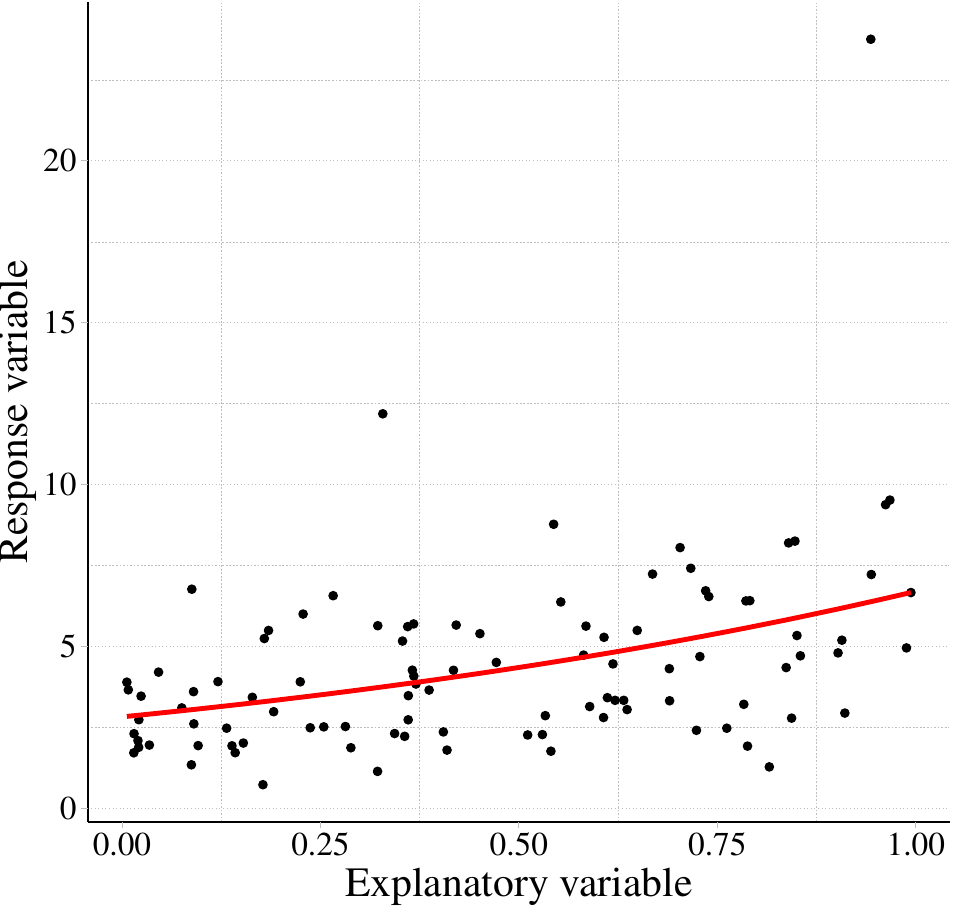}}
		\subfigure[\label{F:qqplot2} Normal Q-Q Plot]{\includegraphics[width=.4\linewidth]{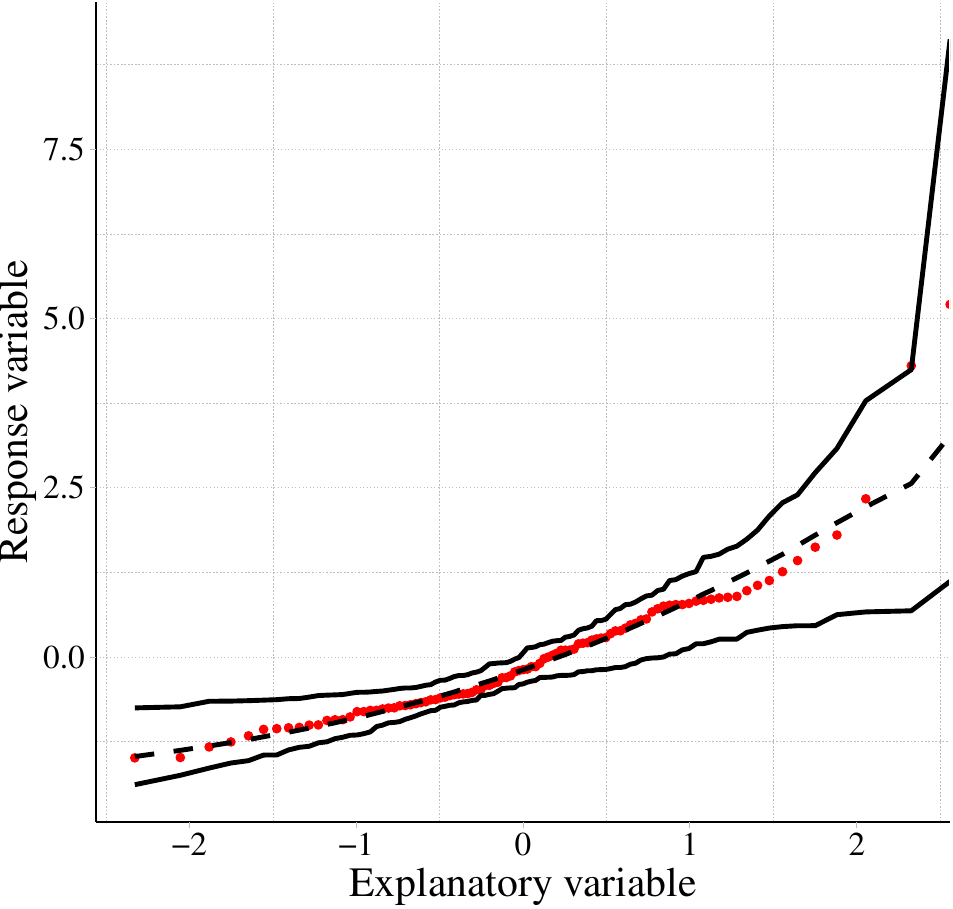}}
		\\
		\subfigure[\label{F:cook2} Cook distance versus predicted values]{\includegraphics[width=.24\linewidth]{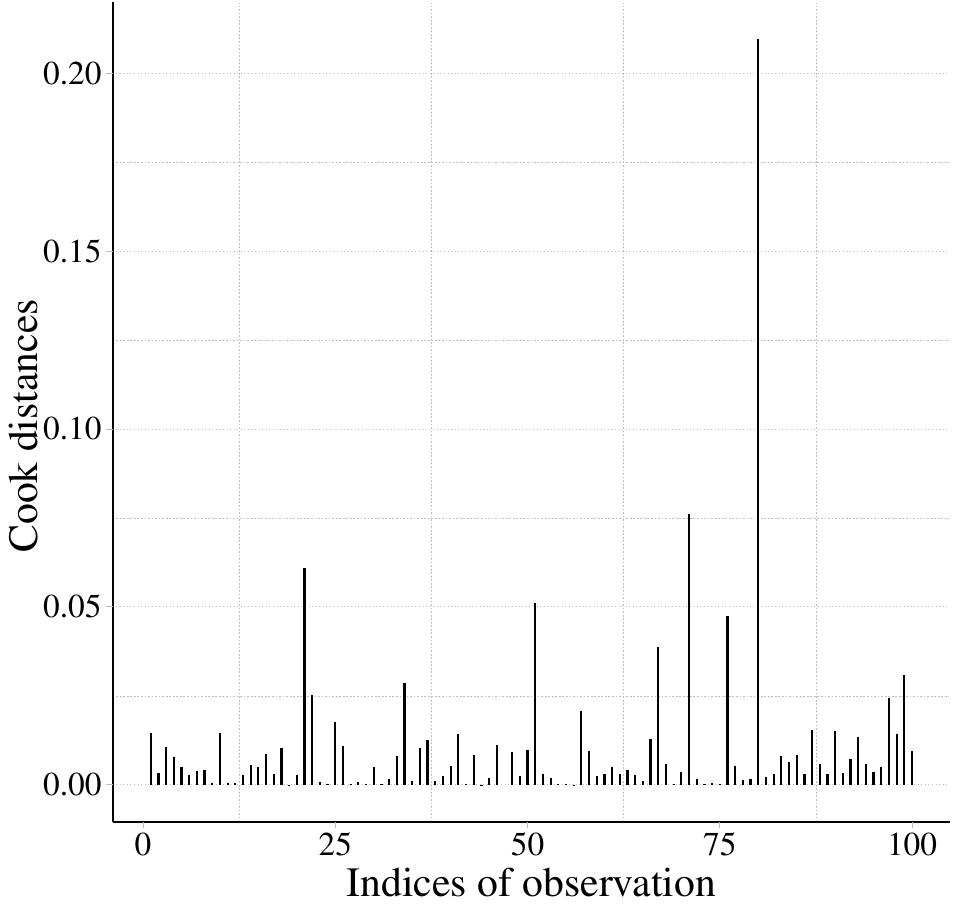}}
		\subfigure[\label{F:hii2} Leverage (h$_{k k}$) versus predicted values]{\includegraphics[width=.24\linewidth]{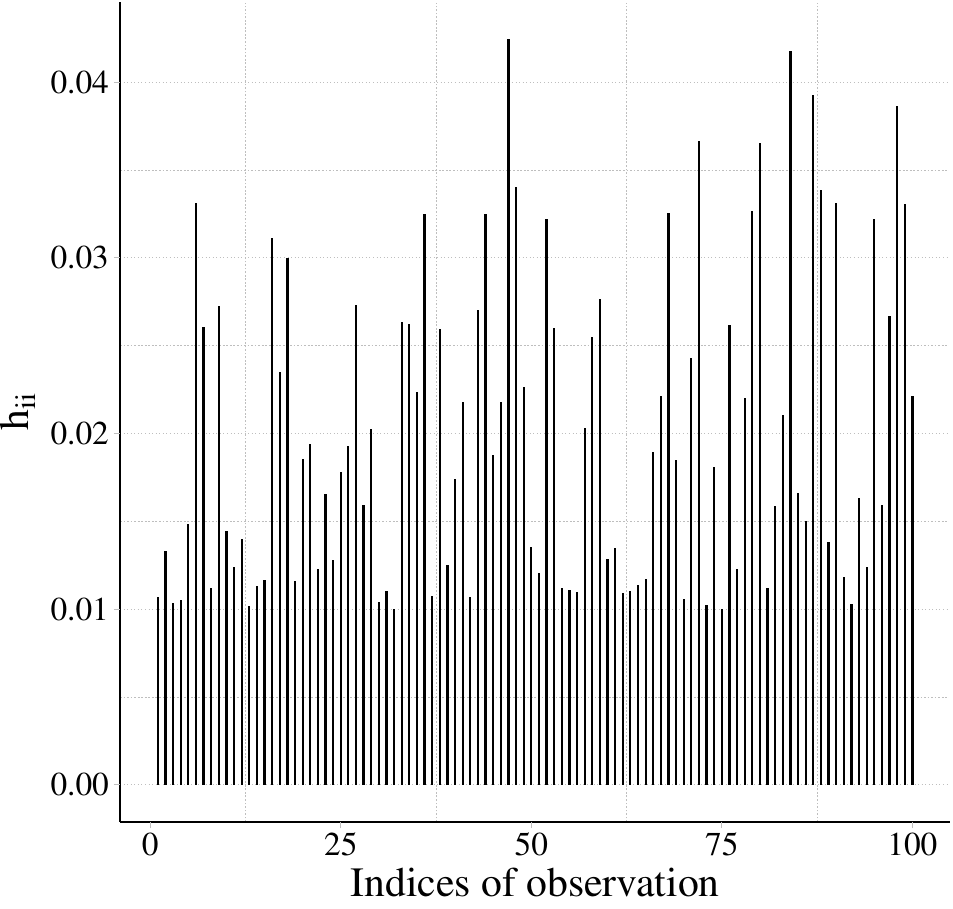}}
		\subfigure[\label{F:dffits1} DFFITS versus indices of observation]{\includegraphics[width=.24\linewidth]{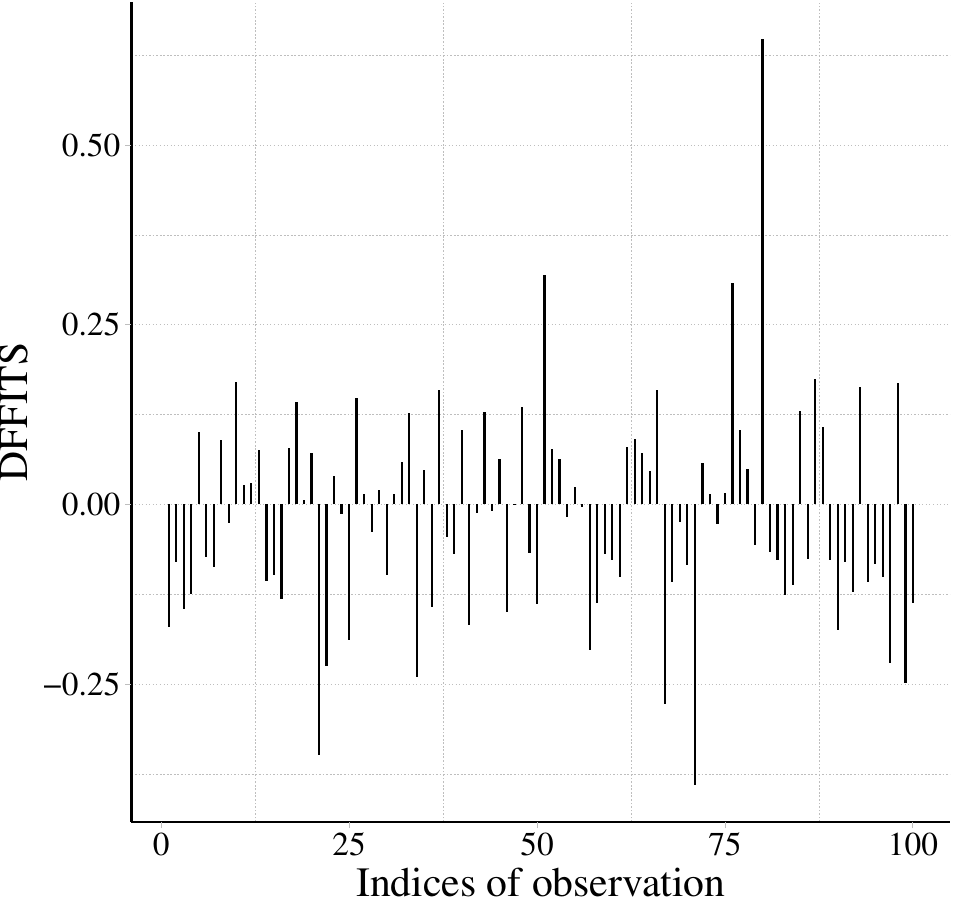}}
		\subfigure[\label{F:res2} Standardized residuals versus predicted values]{\includegraphics[width=.24\linewidth]{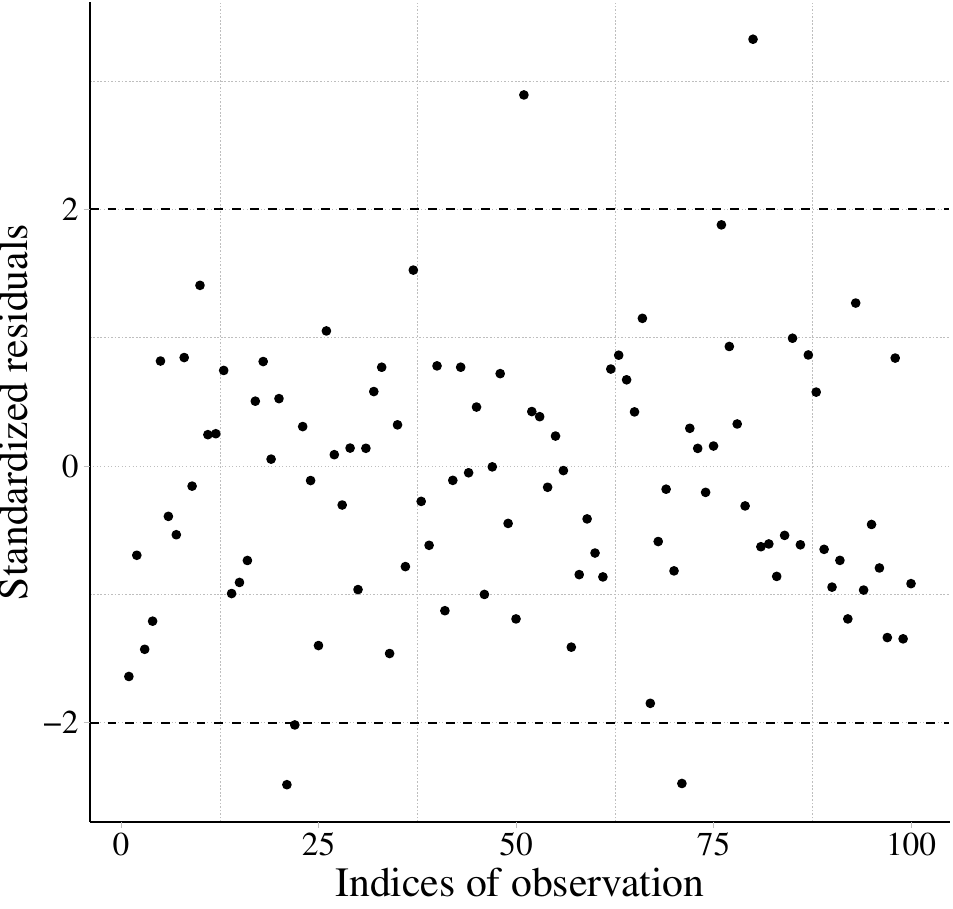}}
		\caption{The upper (a) is a fitted line plot,
			(b), (c) and (d) panels show the
			Cooks distance, leverage, and DFFITS
			against the predicted.
			The (e) panel shows the standardized residuals
			versus predicted and the
			another panel displays a Normal Q-Q plot
			of absolute deviance residuals with a simulated envelope.}
		\label{F:DIAG1}
	\end{figure}

	To illustrate the proposed influence measures, we calculate them for $\G$-distributed data according to the configuration $({\bm \beta}^\top,\alpha,L) = (1,1,-50,4)$ and $n=100$.
	The results are shown in Fig.~\ref{F:DIAG1}.
	As expected, Fig.~\ref{F:estSim} indicates that the model is well-fitted.
	The envelope curve of studentized residuals is shown in Fig.~\ref{F:qqplot2} and confirms the quality of the fit.
	Fig.~\ref{F:cook2} shows the Cook distances compared to the predicted values.
	Observations larger than $\sfrac{8}{(n-2p)}$ are possible influential or aberrant points.
	We also show the values of $(h_{k k})$ versus the predicted values in Fig.~\ref{F:hii2} and, for this measure, observations higher than $\sfrac{3p}{n}$ are possible leverage points.
	The plot of DFFITS against observed values is shown in Fig.~\ref{F:dffits1}.
	Here, observations larger than $2\sqrt{\sfrac{p}{(n-p)}}$ are possible influential or aberrant points.
	Finally, Fig.~\ref{F:res2} shows the plot of studentized residuals versus predicted values.
	There are no visible patterns, as expected, for a good fit.

	\section{A simulation study}\label{sec:SS}

	MLEs for $\G$ regression parameters given in \eqref{E:MLEb}, \eqref{E:MLEa}, and~\eqref{E:MLEl} do not have closed-form expressions and thus iterative numerical methods are required.
	To this end, we conducted a pilot study to select an iterative method.
	In this first study, we selected four methods:
	BFGS (Broyden-Fletcher-Goldfarb-Shanno),
	CG (Conjugate Gradients),
	NM (Nelder-Mead), and SANN (Simulated ANNealing).
	We consider a fixed set of coefficients ${\bm \beta} =(0.01, 0.01, 0.01)$ and number of looks $ L \in\{1,4,8\}$,
	and varying roughness
	$\alpha $ $\in$ $ \{-50, $ $-10, $ $-5, $ $-3\}$,
	and sample sizes $n \in \{20, 50, 100, 500\}$.
	For each combination $(\alpha,n)$ we generate $100$ Monte Carlo replications against which the proposed estimators are evaluated.
	All computational manipulations were performed using the software {\tt R}~\citep{R2015}.
	Using the root mean square error (RMSE) as a metric, CG provided the best results.
	From now on, we will use the method CG.

	Now, we are in a position to conduct a performance study.
	We used 1,000 Monte Carlo replications with observations from random variables following the law
	$\G\big(\alpha,\exp({\beta_0+\beta_1 x_1+\beta_2 x_2}) (-\alpha-1),L\big)$ and the following specifications:
	$n \in \{20, 50, 100, 500\}$,
	$\beta_0=\beta_1=\beta_2 \in \{0.01, 1, 2\}$,
	$\alpha \in \{-15, -10, -5, -3\}$,
	$ L  \in \{1, 4, 8\}$,
	and
	$X_k \sim \G\big(\alpha,(-\alpha-1),L\big)$.

\begin{figure}[!htbp]
		\centering
		\includegraphics[width=\linewidth]{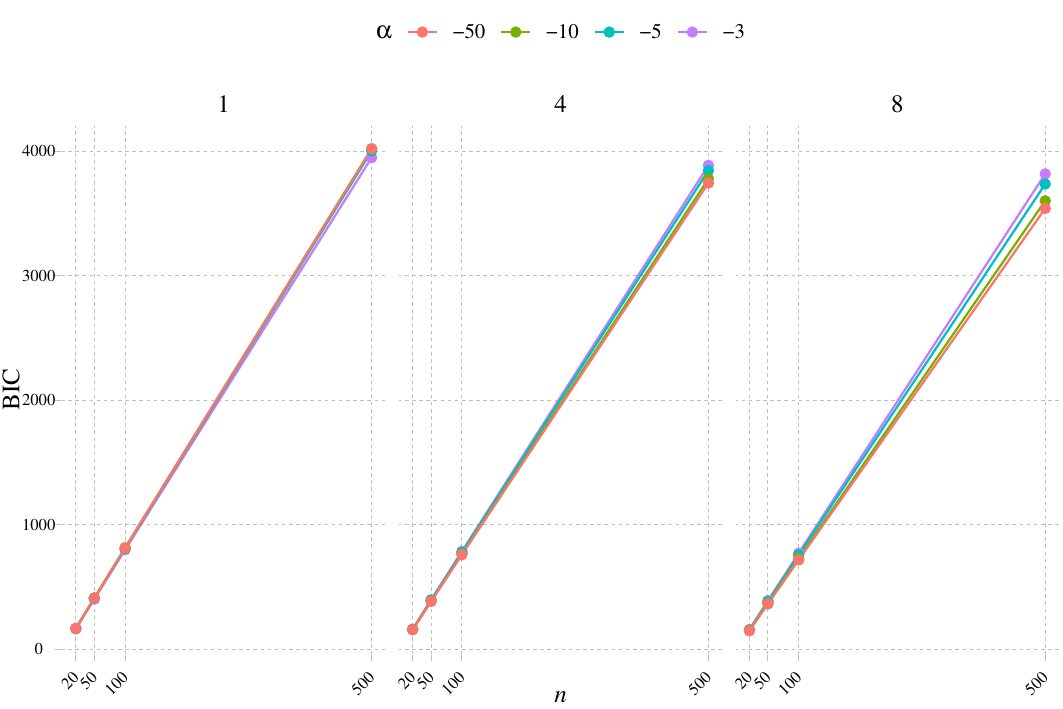}
		\caption{BIC values for samples of size $n \in \{20, 50,  100,  500\}$, $L\in\{1,4,8\}$ looks, and $\alpha\in\{-50,-10,-5,-3\}$.}
		\label{F:infC}
	\end{figure}

	We used five comparison criteria:
	absolute bias (Abias),
	root mean square error (RMSE),
	Akaike information criterion (AIC) and its corrected version AICc,
	and the Bayesian information criterion (BIC).
	The four performance measures behave alike.
	The quality of the estimates improves with the increase of the sample size, as expected; cf.\
	Fig.~\ref{F:infC}, which shows the BIC.

	\section{Application to SAR imagery}
	\label{sec:app}

	\subsection{Comparison study}

	In this section, we apply the proposed $\G$ regression model to the analysis of SAR images and compare the agreement between the fitted and actual values using the other seven models:
	Exponential, Gamma ($\Gamma$), Reciprocal Gamma ($\Gamma^{-1}$), Normal ($\mathcal N$),
	Inverse Normal ($\mathcal N^{-1}$), Weibull, Power Exponential, and EGB2 regression models.
	It is important to note that the same linear predictors are used in the competing regression models to ensure the conditions for unequal comparison.
	We describe intensities due in the HH and VV channels in terms of HV intensities.
	A preliminary descriptive analysis shows that HV data correlate well with HH and VV data: $0.5486$ and $0.4465$, respectively.

\begin{figure}[!htbp]
		\centering
		\includegraphics[width=0.70\linewidth]{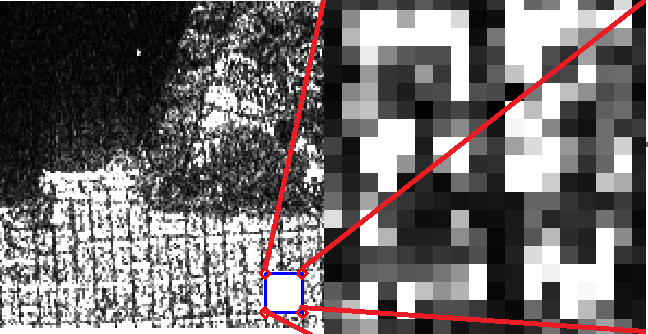}
		\caption{PolSAR image with selected region San Francisco (USA).}
		\label{F:image}
	\end{figure}

	We have used two linear regression models with the same link function:
	(i)~$\mu_{\text{HH}}(x_k) = \exp(\beta_0 + \beta_1 x_k)$, denoted as $(\text{HH} \sim \text{HV})$,
	and
	(ii)~$\mu_{\text{VV}}(x_k) = \exp(\beta_0 + \beta_1 x_k)$, denoted as $(\text{VV} \sim \text{HV})$.
	Below, we describe SAR intensities obtained with the AIRSAR~\citep{LeePottier2009PolarimetricRadarImaging} sensor from scenes in the San Francisco region (USA).
	Fig.~\ref{F:image} shows an intensity HH map of this region.

\begin{figure}[!htbp]
		\centering
		\subfigure{\includegraphics[width=.3\linewidth]{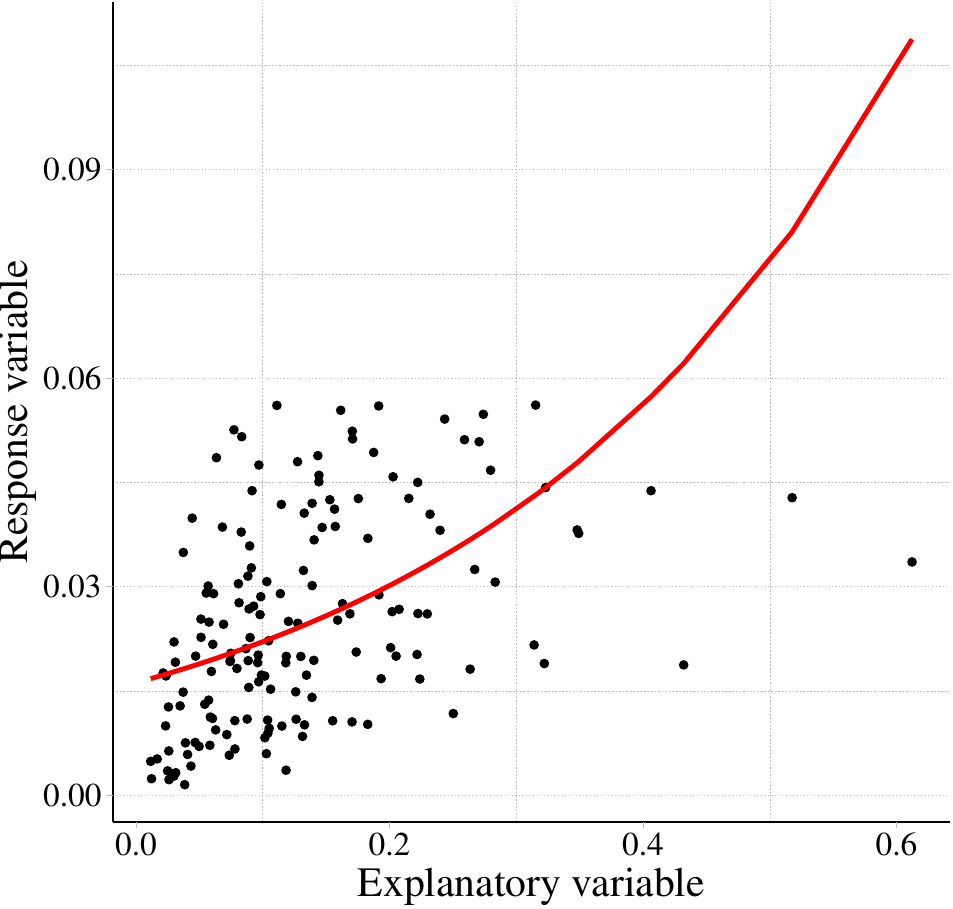}}
		\subfigure{\includegraphics[width=.3\linewidth]{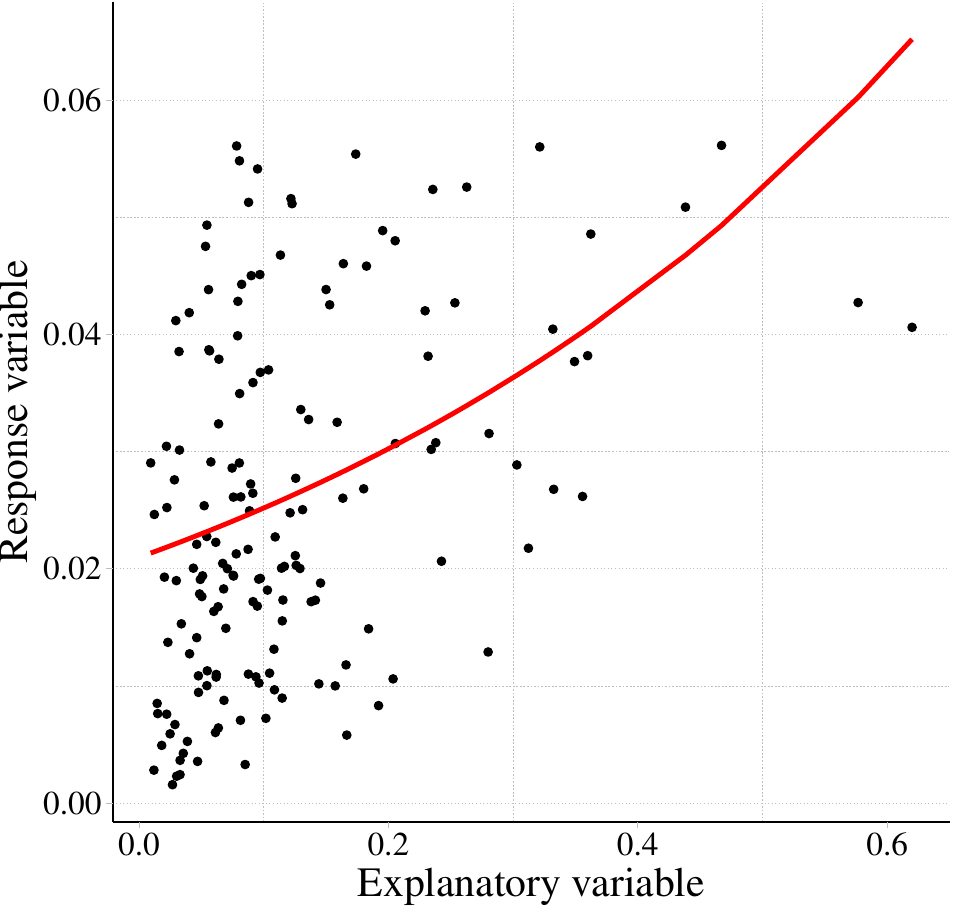}}\\
		\subfigure{\includegraphics[width=.3\linewidth]{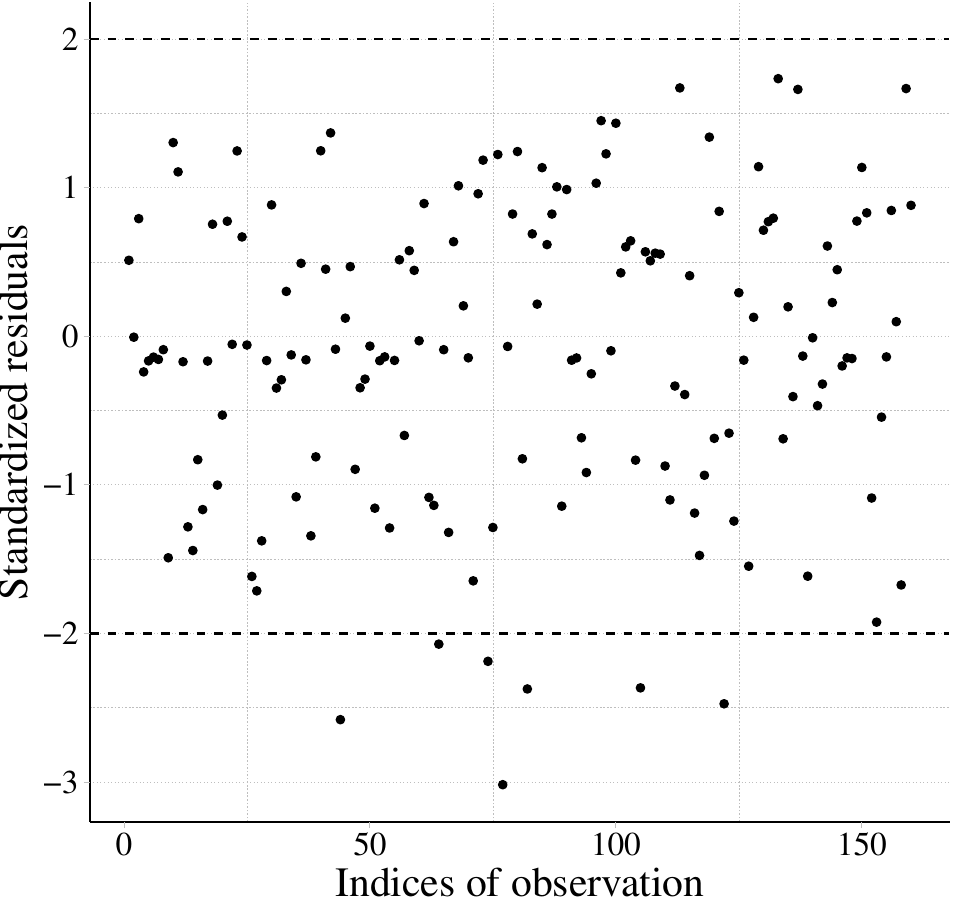}}
		\subfigure{\includegraphics[width=.3\linewidth]{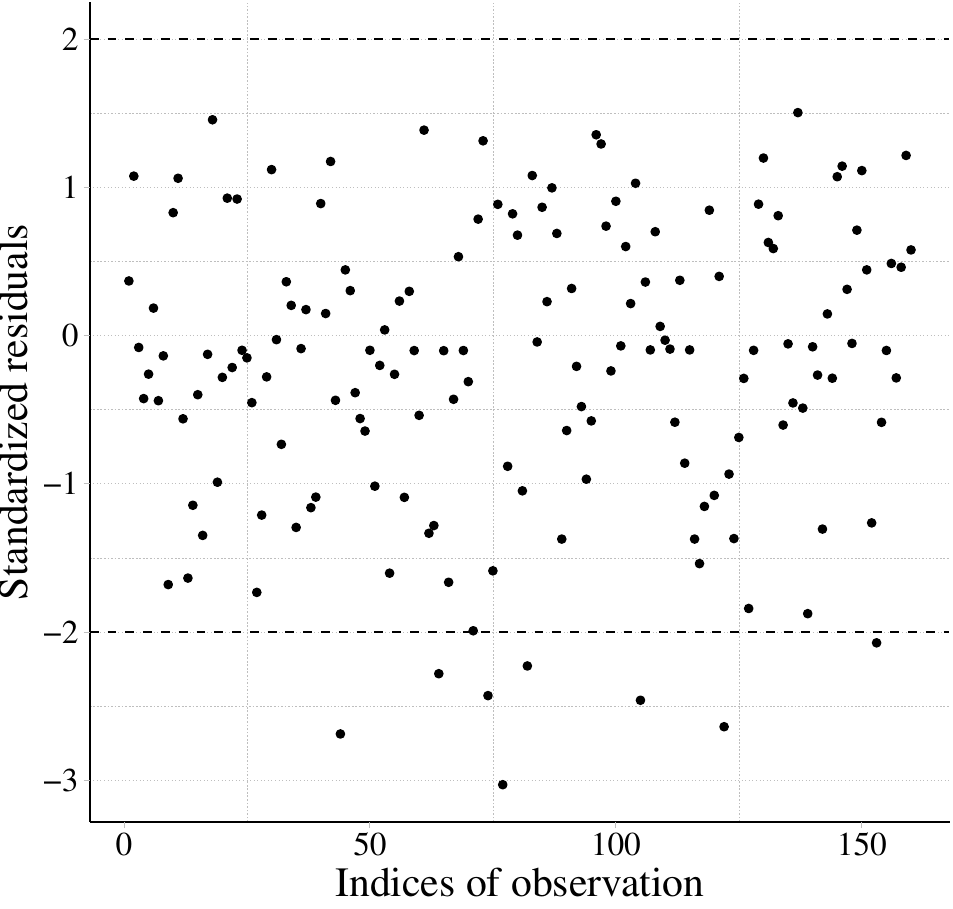}}\\
		\subfigure{\includegraphics[width=.3\linewidth]{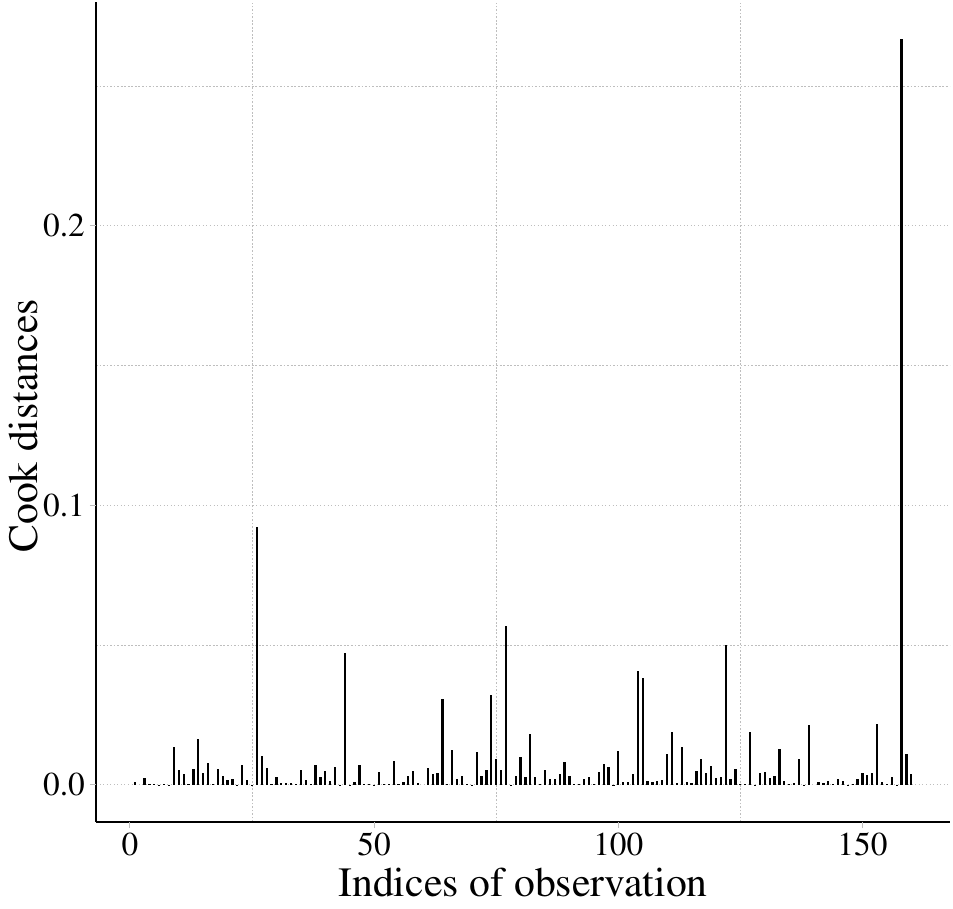}}
		\subfigure{\includegraphics[width=.3\linewidth]{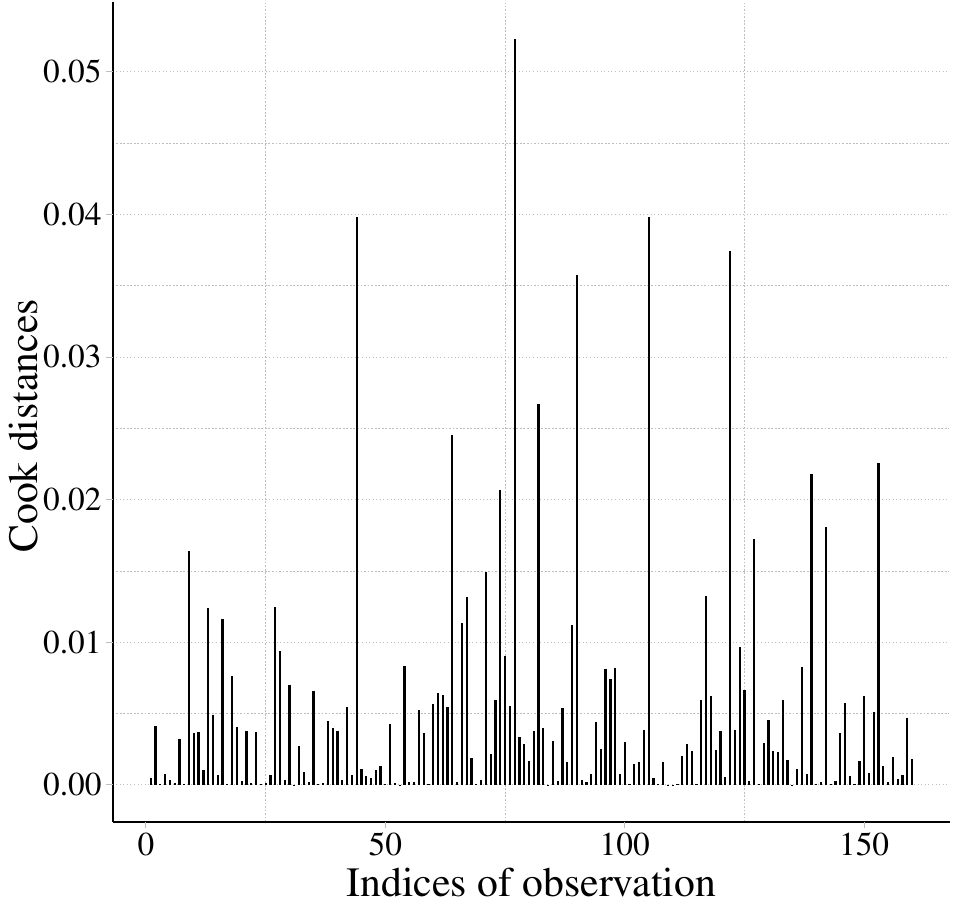}}\\
		\subfigure{\includegraphics[width=.3\linewidth]{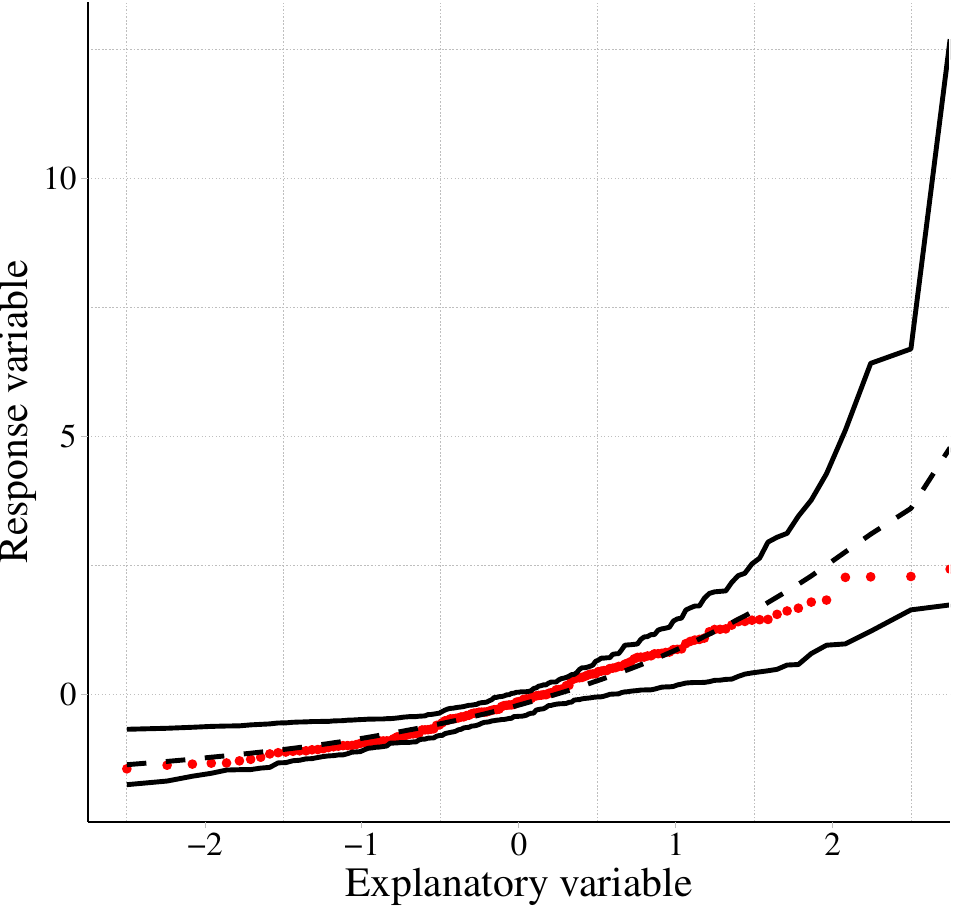}}
		\subfigure{\includegraphics[width=.3\linewidth]{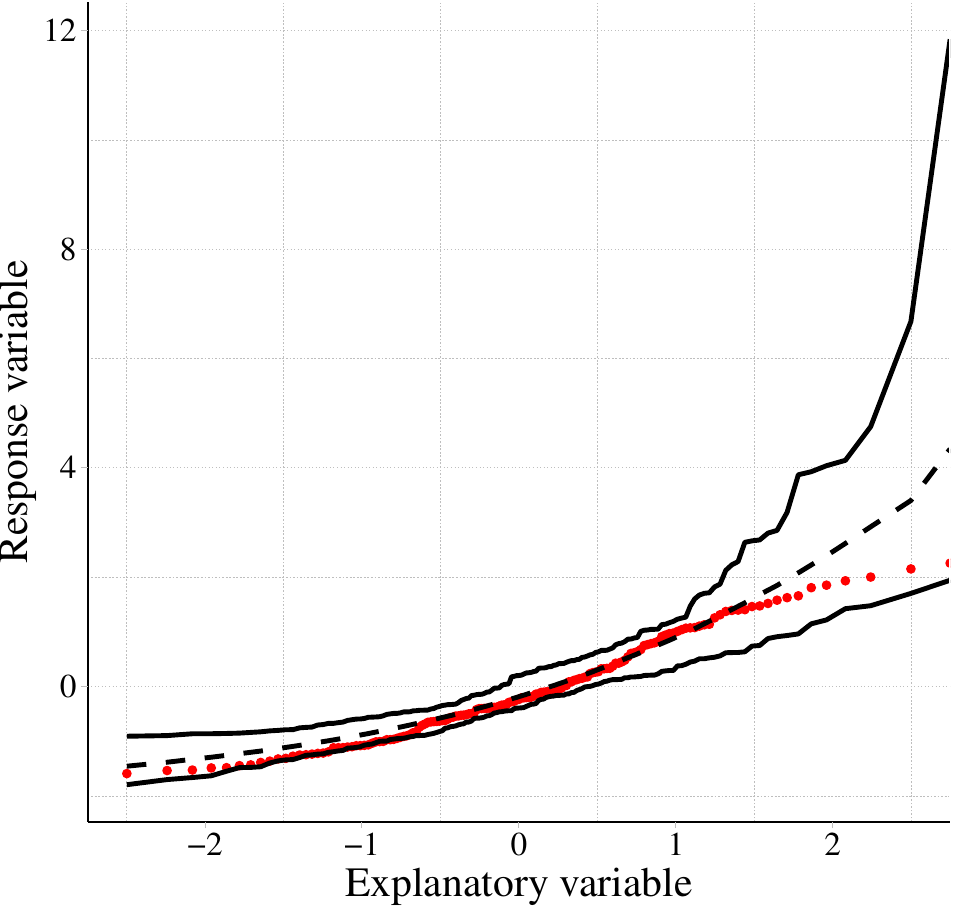}}
		\caption{A fitted line plot and the three diagnostic
			plots for urban SAR imagery data of the two models
			(left panels plots
			are of the model I
			and right band plots
			are of the second model).}
		\label{F:application}
	\end{figure}

	The $\G$ regression model is equipped with the link function
	$$
	\log(\mu_k) = \beta_0 + \beta_1 x_k,
	$$
	where the response variable is $Z_k\sim\G(\alpha, \mu_k(-\alpha-1))$ and the roughness $\alpha$ is a nuisance parameter.
	We use two measures of performance: the mean absolute biases (MAB) and the root mean square error (RMSE), given as:
	$$
	\frac{1}{n}\sum_{k=1}^{n}|Z_k - \widehat{Z}_k|,
	\text{ and }
	\sqrt{\frac{1}{n}\sum_{k=1}^{n}(Z_k - \widehat{Z}_k)^2},
	$$
	respectively.
	Table~\ref{T:measures} shows the values for MAB and RMSE, indicating that the $\G$ regression model overcomes competing models.
	The last panel of Fig.~\ref{F:application} shows the standardized residuals of the considered models.
	Our proposal is the only acceptable model according to this criterion.

	\begin{table}[!htbp]
		\centering
		\caption{Performance measures (MAB, MRAB, and RMSE)
			of the regression models. The best results are in bold.}
		\label{T:measures}
		\begin{tabular}{l rr rr}
			\toprule
			& \multicolumn{2}{c}{$\text{HH} \sim \text{HV}$}   & \multicolumn{2}{c}{$\text{VV} \sim \text{HV}$} \\
			\cmidrule(lr){2-3} \cmidrule(lr){4-5}
			Model              & MAB   & RMSE   & MAB    & RMSE
			\\
			\cmidrule(lr){1-1}
			\cmidrule(lr){2-2}
			\cmidrule(lr){3-3}
			\cmidrule(lr){4-4}
			\cmidrule(lr){5-5}
			$\mathcal{G}^0_I$          & \textbf{0.087} & \textbf{0.126} & \textbf{0.087} & \textbf{0.127} \\
			Exponential                & 0.096 & 0.135 & 0.096 & 0.137 \\
			Gamma                      & 0.096 & 0.135 & 0.096 & 0.137 \\
			Inverse gamma              & 0.104 & 0.161 & 0.102 & 0.164 \\
			Normal                     & 0.096 & 0.135 & 0.097 & 0.137 \\
			Inverse normal             & 0.090 & 0.131 & 0.099 & 0.139 \\
			Weibull                    & 0.092 & 0.139 & 0.091 & 0.139 \\
			Power exponential          & 0.092 & 0.139 & 0.092 & 0.134 \\
			EGB2                       & 0.131 & 0.186 & 0.137 & 0.192 \\
			\bottomrule
		\end{tabular}
	\end{table}

	The MLEs for the parameters involved and their standard errors are shown in Tables~\ref{T:esta} and~\ref{T:estb}.
	All models were well-fitted and showed a significant influence of HV over HH or VV.
	The fitted models are given by
	\begin{align*}
		(\text{HH} \sim \text{HV}): \quad & \widehat{\mu}_{\text{HH}}(x) = \exp(-2.524+17.207x), \\
		(\text{VV} \sim \text{HV}): \quad & \widehat{\mu}_{\text{VV}}(x) = \exp(-2.551+14.608x),
	\end{align*}
	where $x$ is the $\text{HV}$ polarization channel.
	Fig.~\ref{F:application} shows the data with the fitted line, the standardized residuals, the Cook distance, and the normal quantile-quantile plot of the deviance residuals with a simulated envelope~\citep{Atkinson1985}.
	These results show that the new model outpeforms the other competing methods.

	\begin{table}[!htbp]
		\centering
		\caption{Parameter estimates using observed SAR data -- regression $\text{HH} \sim \text{HV}$}
		\label{T:esta}
		\begin{tabular}{lc rrrr}
			\toprule
			Model        & Parameter & Estimate & Std. error & $t$ stat & $p$-value \\
			\cmidrule(lr){1-1}
			\cmidrule(lr){2-2}
			\cmidrule(lr){3-3}
			\cmidrule(lr){4-4}
			\cmidrule(lr){5-5}
			\cmidrule(lr){6-6}
			$\G$              & $\beta_0$ & $-$2.524 & 0.077 & $-$32.800 & 0.000\\
			& $\beta_1$ &   17.207 & 1.613 &    10.700 & 0.000 \\
			Exponential       & $\beta_0$ & $-$2.514 & 0.123 & $-$20.510 & 0.000  \\
			& $\beta_1$ &   16.896 & 2.592 &     6.520 & 0.000   \\
			Gamma             & $\beta_0$ & $-$2.514 & 0.076 & $-$33.000 & 0.000    \\
			& $\beta_1$ &   16.896 & 1.610 &    10.500 & 0.000     \\
			Inverse gamma     & $\beta_0$ & $-$3.344 & 0.079 & $-$42.200 & 0.000\\
			& $\beta_1$ &   17.615 & 1.512 &    11.700 & 0.000 \\
			Normal            & $\beta_0$ & $-$2.414 & 0.108 & $-$22.290 & 0.000  \\
			& $\beta_1$ &   14.879 & 1.548 &     9.610 & 0.000   \\
			Inverse normal    & $\beta_0$ & $-$2.610 & 0.094 & $-$27.900 & 0.000  \\
			& $\beta_1$ &   19.845 & 2.867 &     6.920 & 0.000   \\
			Weibull           & $\beta_0$ & $-$2.385 & 0.078 & $-$30.700 & 0.000    \\
			& $\beta_1$ &   16.592 & 1.619 &    10.200 & 0.000     \\
			Power exponential & $\beta_0$ & $-$2.649 & 0.023 &$-$113.600 & 0.000\\
			& $\beta_1$ &   15.904 & 0.192 &    82.800 & 0.000 \\
			EGB2              & $\beta_0$ & $-$4.265 & 0.557 &  $-$7.660 & 0.000    \\
			& $\beta_1$ &   23.989 & 5.068 &     4.730 & 0.000     \\    \bottomrule
		\end{tabular}
	\end{table}

	\begin{table}[!htbp]
		\centering
		\caption{Parameter estimates using observed SAR data -- regression $\text{VV} \sim \text{HV}$}
		\label{T:estb}
		\begin{tabular}{lc rrrr}
			\toprule
			Model        & Parameter & Estimate & Std. error & $t$ stat & $p$-value \\
			\cmidrule(lr){1-1}
			\cmidrule(lr){2-2}
			\cmidrule(lr){3-3}
			\cmidrule(lr){4-4}
			\cmidrule(lr){5-5}
			\cmidrule(lr){6-6}
			$\G$              & $\beta_0$ & $-$2.551 & 0.100 & $-$25.48 & 0.000\\
			& $\beta_1$ &   14.608 & 1.985 &     7.36 & 0.000 \\
			Exponential       & $\beta_0$ & $-$2.598 & 0.122 & $-$21.26 & 0.000  \\
			& $\beta_1$ &   15.340 & 2.579 &     5.95 & 0.000   \\
			Gamma             & $\beta_0$ & $-$2.598 & 0.087 & $-$29.80 & 0.000    \\
			& $\beta_1$ &   15.340 & 1.841 &     8.33 & 0.000     \\
			Inverse gamma     & $\beta_0$ & $-$3.576 & 0.094 & $-$38.26 & 0.000\\
			& $\beta_1$ &   13.882 & 1.815 &     7.65 & 0.000 \\
			Normal            & $\beta_0$ & $-$2.510 & 0.126 & $-$19.99 & 0.000  \\
			& $\beta_1$ &   13.542 & 1.839 &     7.36 & 0.000   \\
			Inverse normal    & $\beta_0$ & $-$2.670 & 0.105 & $-$25.33 & 0.000  \\
			& $\beta_1$ &   17.501 & 3.065 &     5.71 & 0.000   \\
			Weibull           & $\beta_0$ & $-$2.509 & 0.089 & $-$28.17 & 0.000    \\
			& $\beta_1$ &   15.637 & 1.869 &     8.37 & 0.000     \\
			Power exponential & $\beta_0$ & $-$2.884 & 0.044 & $-$65.00 & 0.000\\
			& $\beta_1$ &   14.828 & 0.739 &    20.10 & 0.000 \\
			EGB2              & $\beta_0$ & $-$7.570 &  2.31 &  $-$3.28 & 0.001    \\
			& $\beta_1$ &   51.760 & 24.25 &     2.13 & 0.034     \\
			\bottomrule
		\end{tabular}
	\end{table}

	\subsection{Predictive study}

	Now we use the Eq.~\eqref{E:GLM} to predict the averaged co-polarized intensity from the cross-polarized intensity.
	An important fact is that the associated error at the $(i,j)$-entry is, in our proposal, $\epsilon(i,j) \sim\mathcal{G}^0_I\big(\alpha,(-\alpha-1),L\big)$ and therefore the associated residue should be marginally distributed as $\mathcal{G}^0_I$-law.
	The initial assumption comes from Equation~\eqref{E:GLM} when formulating the model.
	Accurate determination of the residue distribution is in a difficult direction, as discussed by~\citet{Loynes1969}, but it is possible to raise indications of the adequacy of the model for error.
	In the following, we will analyze the fit and interpretation of our model and discuss the associated error model.

	Among the possible motivations is that reconstruction by regression helps to identify changes, as \citet[section 2.3]{Murat2012}, and this information is important, especially for countries with a sizeable continental extent, such as Brazil.
	We used a dual-pole image acquired by the Sentinel-1 satellite with a spatial range and azimuth resolution \qtyproduct{7x14}{\meter} on April 17, 2023, in the region of Japaratinga, Alagoas, Brazil.
	The system provides values for the HV and VV intensities with $L=3$ looks.
	We used an $11\times 11$ pixels window surrounding each pixel to which we fitted the regression model.
	We will provide: (i)~fit maps $\widehat{\mu}_\text{VV}(z_\text{VV})=\widehat{\mathbbm{E}}({Z}_\text{VV}\mid Z_\text{VV}=z_\text{VV})=\widehat\beta_0+\widehat\beta_1 z_\text{VV}$, (ii)~intercept and slope maps, (iii)~ratio maps between observed and predicted returns to VV intensity, and (iv)~an analysis of the ratio map adherence to the assumed error.

\begin{figure}[!htbp]
		\centering
		\subfigure[\label{Ima:Japa0} Optical image]{\includegraphics[width=.28\linewidth]{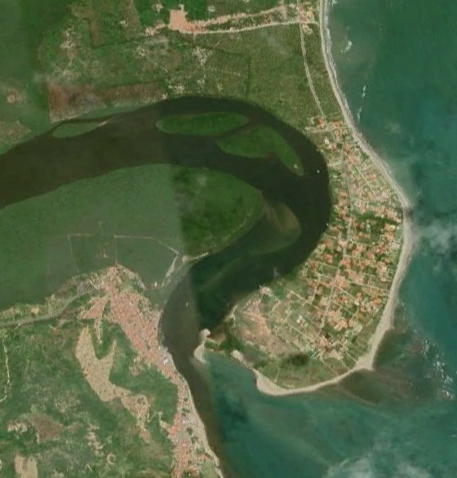}}
		\subfigure[\label{Ima:Japa12} VV intensity image]{\includegraphics[width=.3\linewidth]{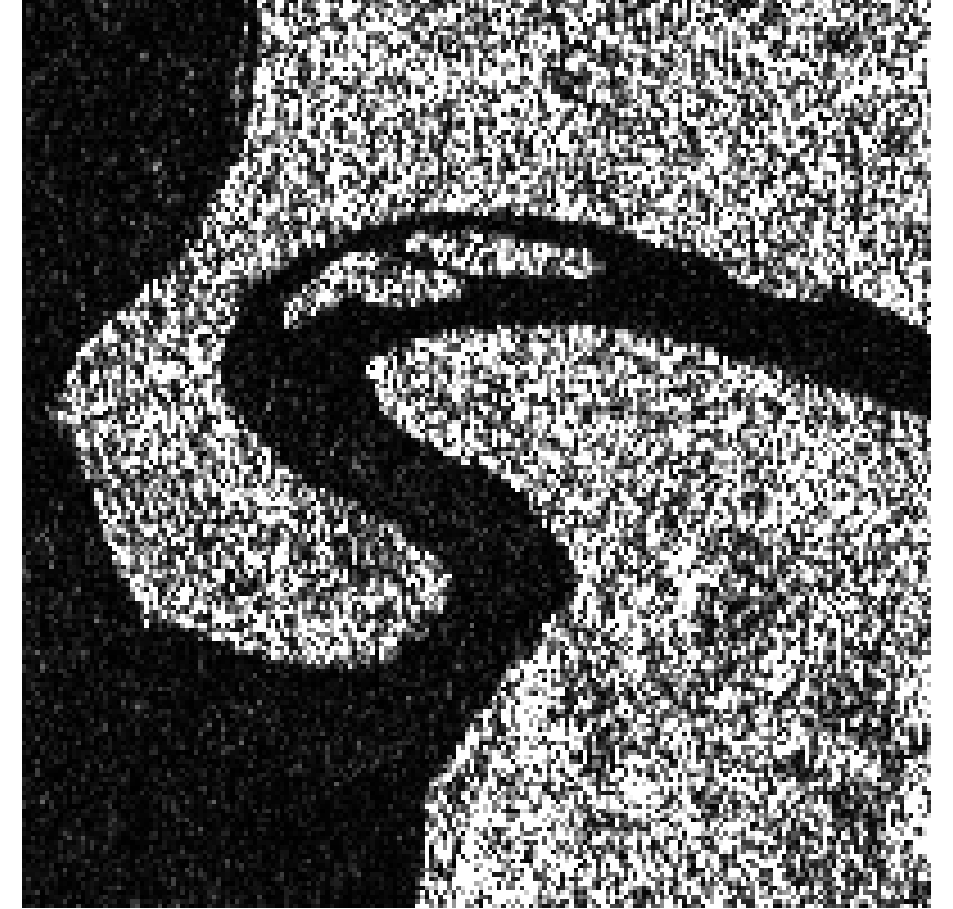}}
		\subfigure[\label{Ima:Japa2} Predicted image, $\widehat{\mu}_\text{VV}(z_\text{VV})=\widehat{\mathbbm{E}}({Z}_\text{VV}\mid Z_\text{VV}=z_\text{VV})=\widehat\beta_0+\widehat\beta_1 z_\text{VV}$]{\includegraphics[width=.3\linewidth]{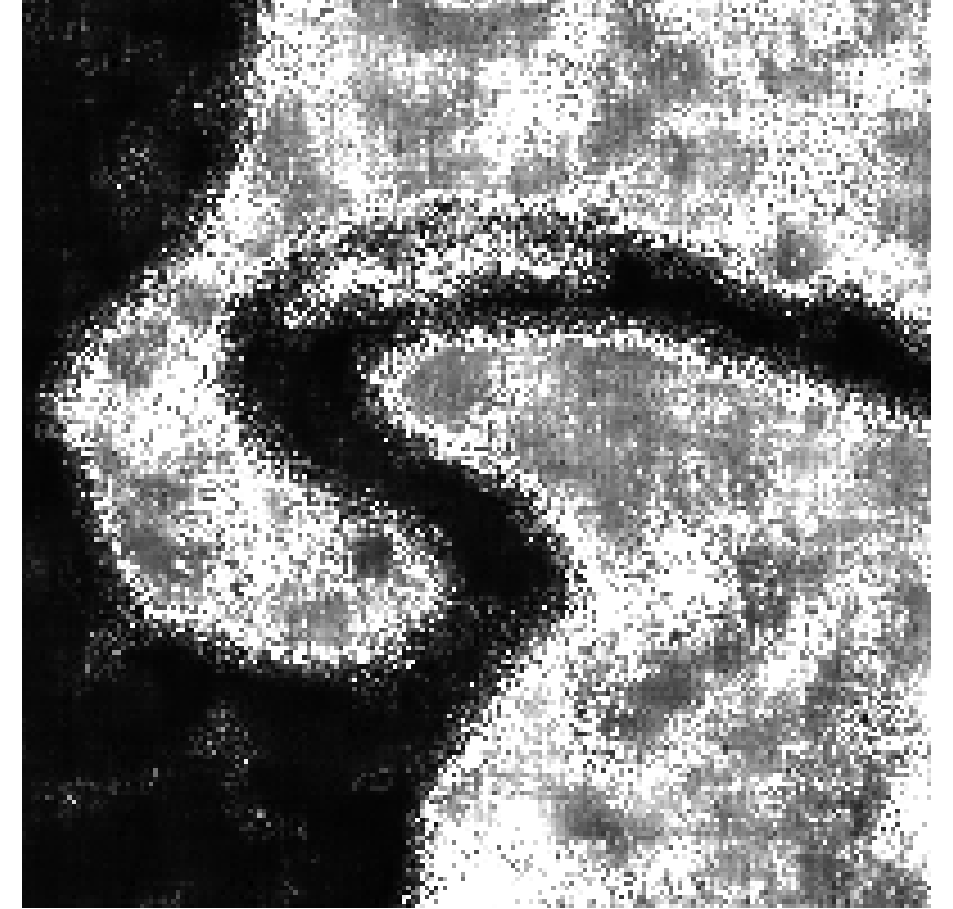}}\\
		\subfigure[\label{Ima:Japa3} Intercept image, $\widehat\beta_0$]{\includegraphics[width=.3\linewidth]{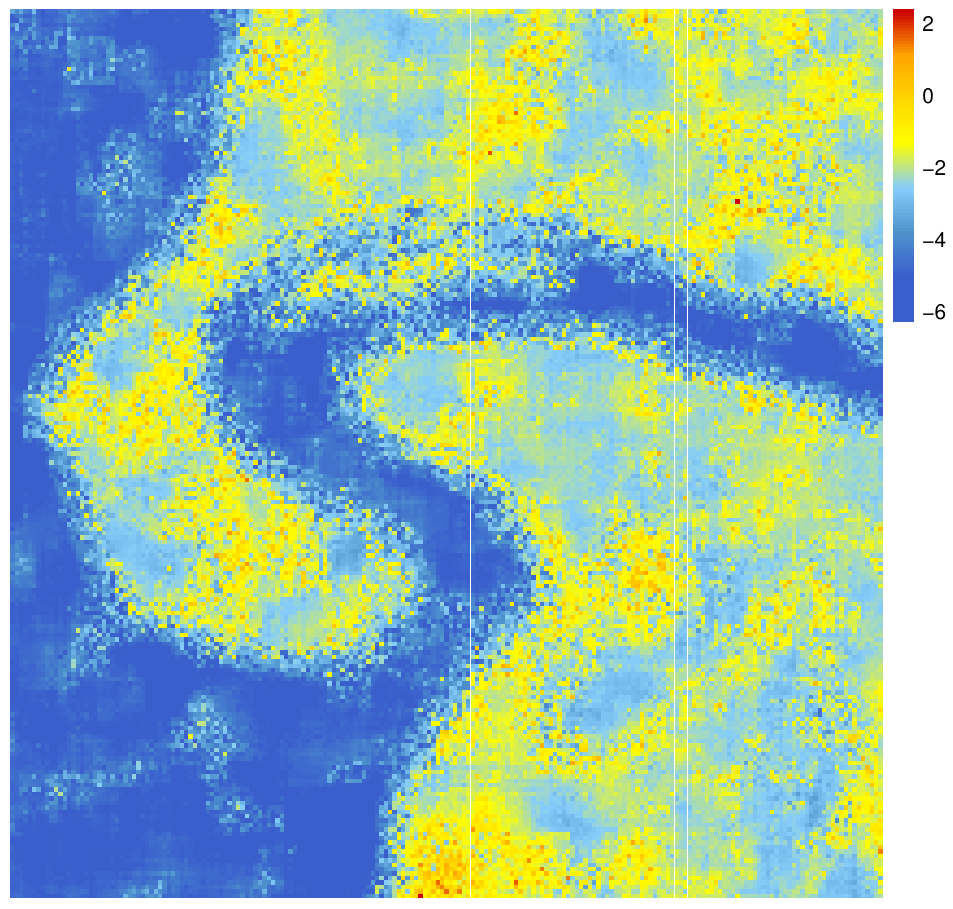}}
		\subfigure[\label{Ima:Japa4} Slope image, $\widehat\beta_1$]{\includegraphics[width=.3\linewidth]{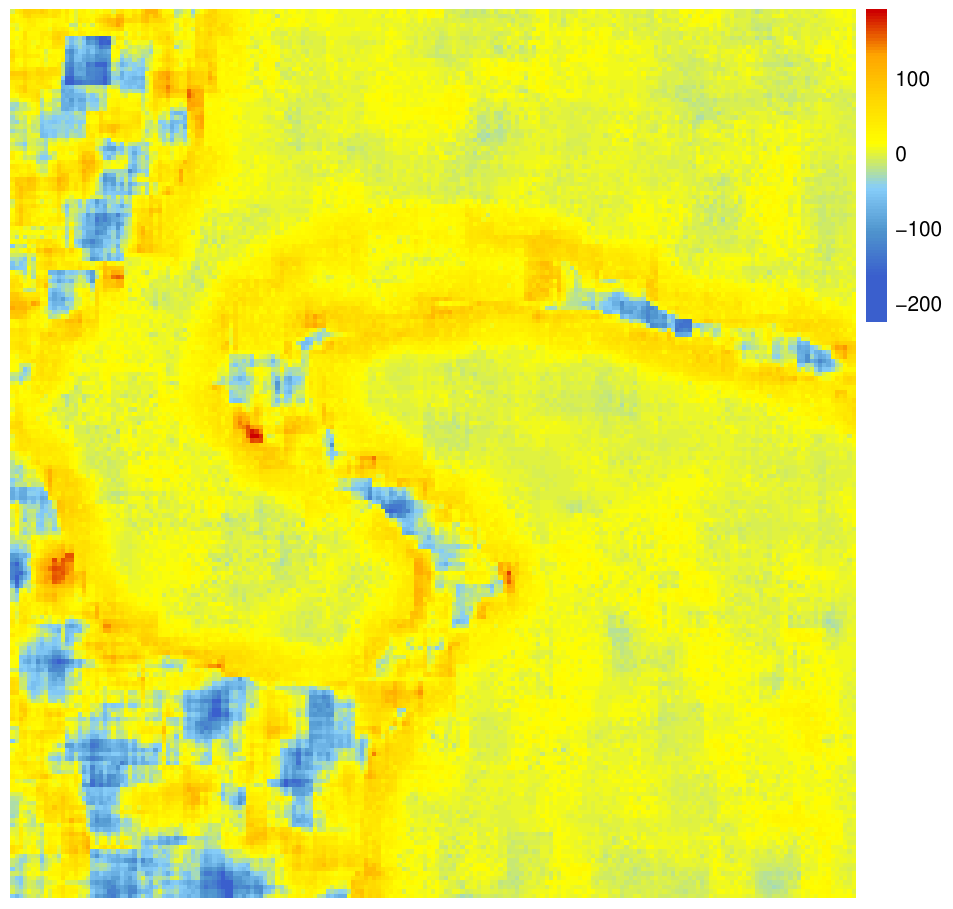}}\\
		\subfigure[\label{Ima:Japa5} Ratio image, ${Z}_\text{VV}/\widehat{\mu}_\text{VV}$]{\includegraphics[width=.3\linewidth]{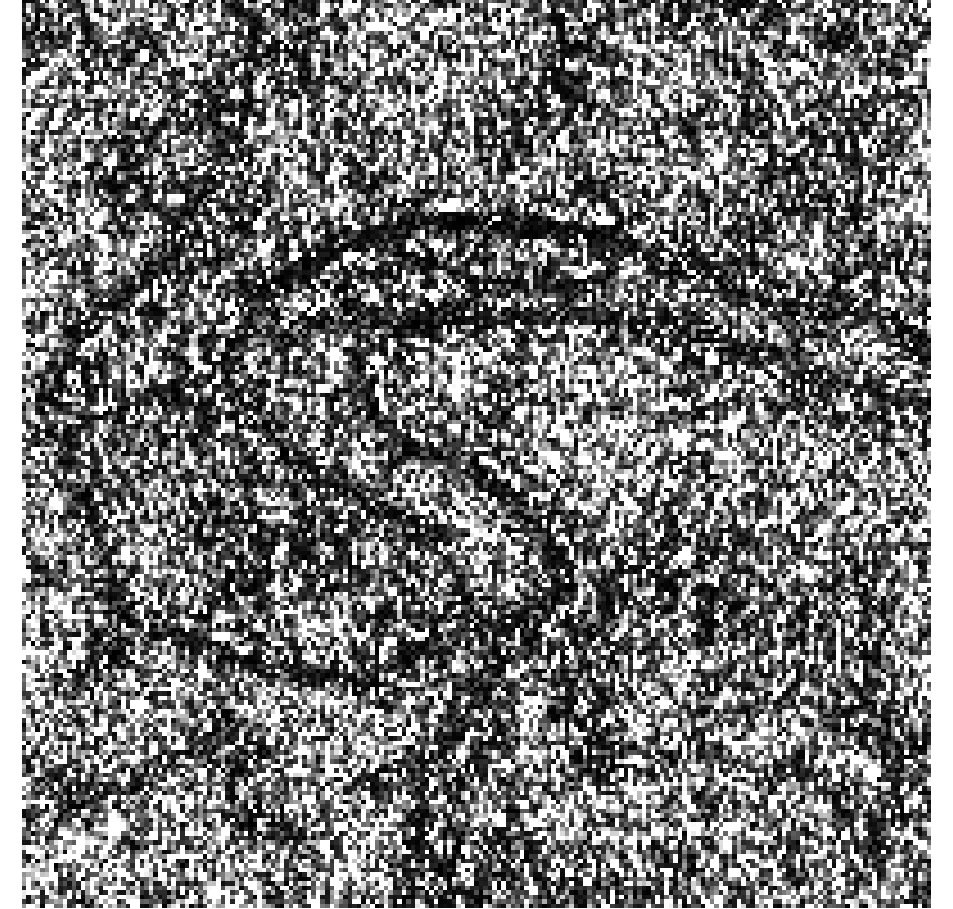}}
		\subfigure[\label{Ima:Japa6} Empirical and theoretical cdf]{\includegraphics[width=.28\linewidth]{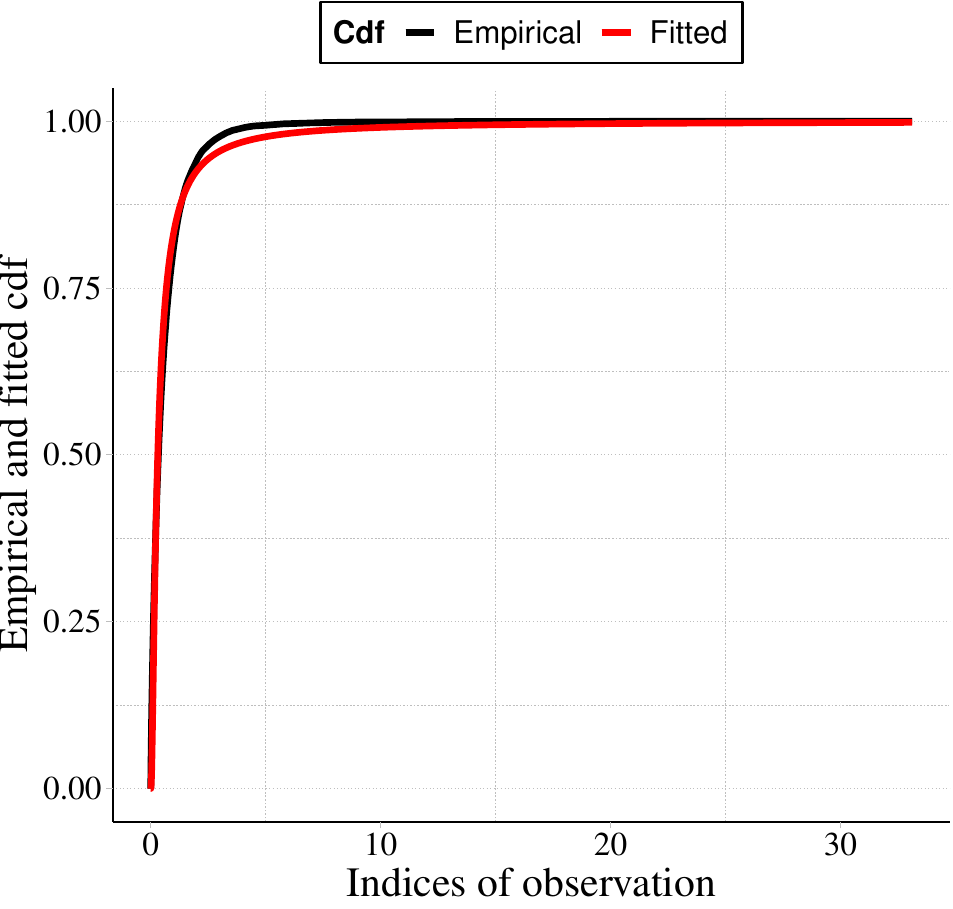}}
		\caption{Exploratory analysis on predictions and residuals.}
		\label{Ima:Japa}
	\end{figure}

	Figure~\ref{Ima:Japa0} displays the optical image, showing a lake, forest, and built-up area regions.
	The observed and predicted images (using the HV intensities as explanatory variables) of VV intensities are shown in Figures~\ref{Ima:Japa12} and~\ref{Ima:Japa2}, respectively.
	Despite using the simplest possible model, the reconstructed image detects texture changes.
	Another possibility is to use feature matrices (such as spatial information from parametric and non-parametric methods) or design of experiments.
	These two ways will require minor changes to our proposal.

	Figures~\ref{Ima:Japa3} and~\ref{Ima:Japa4} detail the layers of the predicted image and offer an understanding of the effect of the cros-polarized channel on the co-polarized one.
	It can be seen from the predicted slope image (Fig.~\ref{Ima:Japa4}) that the HV intensities from the lake regions tend to be inversely proportional in average to the VV intensities.
	At the same time, this relationship is directly proportional for forest and built-up regions.
	Apart from the apparent differences between terrains, the predicted intercept image (Fig.~\ref{Ima:Japa3}) appears to contain subgroups within terrains that are attractive for inputting information into clustering or classification methods of SAR data.

	The ratio image in Figure~\ref{Ima:Japa5} shows a pattern without structure, as expected from the assumption~\eqref{E:GLM}.
	Figure~\ref{Ima:Japa6} confirms what was expected: from the comparison between empirical and $\mathcal{G}^0_I$ cumulative distribution functions (cdf), it appears that the standard $\mathcal{G}^0_I$ model is a good alternative to the ratio data.
	We apply the Cramer-von Mises goodness test to the null hypothesis $\mathcal{H}_0:\text{data come from } \mathcal{G}^0_I(\alpha_0,(-\alpha_0-1),3)$, where $\alpha_0=-1.432434$ (obtained minimum mean square error estimates).
	The $p$-value \SI{4}{\percent} shows that $\mathcal{H}_0$ is not rejected assuming a nominal level of less than 4\%, i.e., the regression model used is not inappropriate.

	It is interesting to note that in Table~\ref{T:estb}, where the null hypothesis is that the regression is not significant, we have obtained lower values indicating its rejection.
	We conclude that our model is commendable in terms of both significance and adherence to the stochastic hypothesis.

	\section{Conclusion}
	\label{sec:conc}

	In this work, we proposed a new $\G$ regression model to describe the conditional SAR intensities.
	The proposed model has been shown to be very flexible for different SAR textures compared to exponential, $\Gamma$, $\Gamma^{-1}$, $\mathcal{N}$, $\mathcal{N}^{-1}$, Weibull, power exponential, and EGB2 regression models.
	We used the maximum likelihood procedure for $\G$ regression parameters and proposed a closed-form expression for the Fisher information matrix.
	A Monte Carlo simulation study evaluated the performance of the estimates.
	Then, some diagnostic and influential techniques were proposed: generalized leverage and Cook measure and two types of residuals.
	Finally, two applications of the $\G$ regression model to actual data from SAR were performed.
	The results showed that our proposal could be useful for processing SAR images with different resolutions: speckled data from the San Francisco (with a high resolution of \qtyproduct{10x10}{\meter}) and Japaratinga (with an average resolution of \qtyproduct{7x14}{\meter}) regions.

	\appendix

	\renewcommand{\thesection}{\Alph{section}}
	\numberwithin{equation}{section}

	\section{Proof of Corollary \ref{C:EV_T}}\label{App2}

	If
	$Z \sim \G(\alpha,\gamma,L)$ and $T=\gamma+L Z$, then,
	by definition of the cdf, we have:

	\begin{equation}\label{E:fdpT}
		F_T(t) = \Pr(T \leq t) = \Pr\Big(Z \leq \frac{t-\gamma}{ L }\Big)=
		F_Z\Big(\frac{t-\gamma}{ L }\Big),
	\end{equation}
	By taking the derivative of both sides the Expression \eqref{E:fdpT} with respect to variable $t$, we have:
	$$
	f_T(t) = \frac{1}{L}f_Z\left(\frac{t-\gamma}{L}\right)
	=
	\frac{\Gamma(L-\alpha)}
	{\Gamma(-\alpha)\Gamma(L)}
	\frac{t^{\alpha-1}}{\gamma^\alpha}
	\Big[1-\frac{\gamma}{t}\Big]^{L-1},
	$$
	where $t \in (\gamma, \infty)$.
	A random variable $1/T$ occurs frequently in the analytic manipulations we use to derive the results of this work.
	Now we are ready to find its $n$th non-central moment:
	\begin{align*}
		\E\Big[\frac{1}{T^n}\Big] &= \int_{\gamma}^{\infty}t^{-n}f_T(t)\operatorname{d}t\\
		&=\int_{\gamma}^{\infty}\frac{\Gamma( L -\alpha)}
		{\Gamma(-\alpha)\Gamma( L )}\frac{t^{\alpha-n-1}}{\gamma^\alpha}
		\Big[1-\frac{\gamma}{t}\Big]^{ L -1}\operatorname{d}t.
	\end{align*}
	With the change of variable $s=\sfrac{\gamma}{t}$ we have:
	$$
	\E\Big[\frac{1}{T^k}\Big]
	=
	\frac{\Gamma(L-\alpha)}{\Gamma(-\alpha)\Gamma(L)}\frac{1}{\gamma^k}
	\int_{0}^{1}s^{-\alpha+k-1}
	\left(1-s\right)^{ L -1}\mathrm{d}s.
	$$
	From simple algebraic manipulations with the beta density, it follows that
	$$
	\E\Big[\frac{1}{T^n}\Big]
	= \frac{1}{\gamma^n}
	\frac{B(-\alpha+n,L)}{B(-\alpha,L)}
	=\frac{1}{\gamma^n}
	\prod_{k=0}^{n-1}
	\Big(\frac{-\alpha+k}{-\alpha+L+ k}\Big).
	$$

	\section{Proof of Lemma \ref{L:SF}}\label{App3}

	The following derivation proves Lemma \ref{L:SF}.
	Let $Z \sim \G(\alpha, \gamma,  L )$ and $c$ a non-negative constant.
	Then if $Z_1 = cZ$, we have from simple algebraic manipulations:
	\begin{equation}
		\label{E:pdfY}
		F_{Z_1}(z_1) = \Pr(Z \leq \sfrac{z_1}{c}),
		\text{ and, then, }
		f_{Z_1}(z_1) = \frac{1}{c}f_{Z}(\sfrac{z_1}{c}).
	\end{equation}
	Thus, using \eqref{E:pdf} in \eqref{E:pdfY} and rearranging the expression, we have:
	$$
	f_{Z_1}(z_1) = \frac{L^L\Gamma(L-\alpha)}
	{(\gamma c)^\alpha\Gamma(-\alpha)\Gamma(L)}{z_1}^{L-1}(\gamma c+ L z_1)^{\alpha-L},
	$$
	i.e. $Z_1 \sim \G(\alpha,\gamma c,L)$.
	Therefore, the $\G$ distribution is a member of a scale family.

	\section{Fisher information matrix and its inverse}\label{App4}

	Now we derive the score function and the Fisher information matrix for the parameter vector ${\bm \theta}$.
	From Expression \eqref{E:LL}, see \citet{Searle1997} for details, the following identity holds:
	\begin{align}\label{E:SF_B}
		\begin{split}
			U_{\bm \beta} &=
			\sum_{k=1}^{n}\frac{\partial \ell_k({\bm \theta})}{\partial {\bm \beta}} =
			\sum_{k=1}^{n} \frac{\partial \ell_k({\bm \theta})}{\partial \mu_k}
			\frac{\partial \mu_k}{\partial\eta_k}
			\frac{\partial\eta_k}{\partial {\bm \beta}}
			\\
			&= \alpha\sum_{k=1}^{n}
			\Big[\frac{(\alpha- L )(-\alpha-1)}{T_k} -
			\frac{1}{\mu_k}\Big]
			\frac{1}{g^\prime(\mu_k)}{\bm x}_{k}.
		\end{split}
	\end{align}
	Next we obtain the matrix expression for the score function for ${\bm \beta}$ that is given in Equation \eqref{E:MLEb}.
	For the parameter $\alpha$, we obtain:
	\begin{align}\label{E:SF_A}
		\begin{split}
			U_{\alpha} %
			= \sum_{k=1}^n\frac{\partial \ell_k({\bm \theta})}{\partial \alpha}
			=n U_1(\alpha,L)+\sum_{k=1}^n\log\frac{T_k}{\mu_k}
			-(\alpha-L)\sum_{k=1}^n\frac{\mu_k}{T_k}.
		\end{split}
	\end{align}
	Similarly, it can be shown that the score function for $L$ can be written as:
	\begin{equation}\label{E:UL}
		U_L = n U_2(\alpha,L) + \sum_{k=1}^n\log\frac{z_k}{T_k}+(\alpha-L)\sum_{k=1}^{n}\frac{z_k}{T_k}.
	\end{equation}
	From regularity conditions, it is known that the expected value of the derivative in Equation \eqref{E:LL} equals zero.

	From Expression \eqref{E:SF_B}, the Hessian function
	in ${\bm \beta}$ is
	\begin{align}\label{E:DBB}
		\begin{split}
			U_{{\bm \beta} {\bm \beta}} =
			\frac{\partial^2 \ell({\bm \theta})}
			{\partial {\bm \beta} \partial {\bm \beta}^\top}
			=\sum_{k=1}^{n}
			\Big[\frac{\partial^2 \ell_k({\bm \theta})}{\partial \mu_k^2}
			\Big(\frac{\partial \mu_k}{\partial \eta_k}\Big)^2
			+
			\frac{\partial \ell_k({\bm \theta})}{\partial \mu_k}
			\frac{\partial^2 \mu_k}{\partial \eta_k^2}
			\Big]
			\bm{x}_k\bm{x}_k^\top.
		\end{split}
	\end{align}
	Since
	$\E({\partial\ell_k({\bm \theta})}/{\partial \mu_k})=0$, we keep
	$$
	\E(U_{{\bm \beta} {\bm \beta}}) = \sum_{k=1}^{n}
	\E\Big[\frac{\partial^2 \ell_k({\bm \theta})}{\partial \mu_k^2}
	\Big]
	\left(\frac{\partial \mu_k}{\partial \eta_k}\right)^2
	\bm{x}_k\bm{x}_k^\top.
	$$
	Using the Expression \eqref{E:wt}, we have
	$$
	\E(U_{{\bm \beta} {\bm \beta}}) = \sum_{k=1}^{n} \Big[
	\frac{\alpha}{\mu_k^2}
	+ c_1 (-\alpha-1)
	\E\Big(\frac{1}{T_k^2}\Big)\Big]
	\frac{\bm{x}_k\bm{x}_k^\top}{g^\prime(\mu_k)^2},
	$$
	and from Corollary \ref{C:EV_T} we have
	$$
	\E\Big(\frac{1}{T_k^2}\Big) =
	\frac{1}{\mu_k^2(-\alpha-1)}
	\frac{\alpha(\alpha-1)}{(L-\alpha+1)c_1}
	$$
	and
	$$
	\E(U_{{\bm \beta}{\bm \beta}}) =
	\alpha\Big(\frac{L}{L-\alpha+1}\Big)
	\sum_{k=1}^n
	\frac{1}{\mu_k^2}\frac{1}{g^\prime(\mu_k)^2}
	\bm{x}_k\bm{x}_k^\top,
	$$
	see in Equation \eqref{E:FIM} the matrix form.
	From Expression \eqref{E:SF_B}, the Hessian function at the ${\bm \beta}$ and $\alpha$ can be written as
	\begin{align*}
		U_{{\bm \beta}^\top\alpha} =
		\frac{\partial [U_{\bm \beta^\top}]}{\partial \alpha}
		= \sum_{k=1}^n
		\Big[
		\frac{(\alpha-L)(-\alpha-1)}{T_k^2}
		- \frac{2\alpha+1- L }{T_k\mu_k}
		-\frac{1}{\mu_k^2}
		\Big]
		\frac{\mu_k {\bm x}_k^\top}{g^\prime(\mu_k)}.
	\end{align*}
	Hence, by applying the expected value, we obtain:
	\begin{align*}
		\E(U_{{\bm \beta}^\top\alpha}) =
		\sum_{k=1}^{n}\Big[
		(\alpha- L )(-\alpha-1)\E\left(\frac{1}{T_k^2}\right)
		- \frac{2\alpha+1- L }{\mu_k}
		\E\Big(\frac{1}{T_k}\Big)
		-\frac{1}{\mu_k^2}
		\Big]\frac{\mu_k {\bm x}_k^\top}{g^\prime(\mu_k)}.
	\end{align*}
	By using the Corollary \ref{C:EV_T}, we have:
	$$
	\E(U_{{\bm \beta}^\top\alpha}) = c_2
	\sum_{k=1}^{n}\frac{1}{g^\prime(\mu_k)}
	\frac{1}{\mu_k}{\bm x}_{k}^\top,
	$$
	where
	$$
	c_2 = -\Big[1+\frac{(2\alpha+1- L )\alpha}{(-\alpha-1)(\alpha- L )}
	+\frac{\alpha(\alpha-1)}{(-\alpha-1)( L -\alpha+1)}\Big],
	$$
	The matrix expression is in Equation \eqref{E:FIM}.
	From Expression \eqref{E:SF_B}, the Hessian function at the ${\bm \beta}$
	and $L$ can be written as
	\begin{equation*}
		U_{{\bm \beta}^\top L } = \frac{\partial [U_{\bm \beta^\top}]}{\partial  L }
		=
		-(-\alpha-1)
		\sum_{k=1}^{n}
		\Big[
		\frac{T_k + (\alpha-L)z_k}{T_k^2}
		\Big]
		\frac{1}{g^\prime(\mu_k)} {\bm x}_{k}^\top.
	\end{equation*}
	Applying the expected value, we get:
	\begin{equation*}
		\E(U_{{\bm \beta^\top} L}) =
		-(-\alpha-1)\sum_{k=1}^n
		\Big[\E\Big(\frac{1}{T_k}\Big) +
		(\alpha-L)\E\Big(\frac{Z_k}{T_k^2}\Big)\Big]
		\frac{1}{g^\prime(\mu_k)}{\bm x}_{k}^\top.
	\end{equation*}
	We derived that
	\begin{equation}
		\label{E:EVBL}
		\E\Big[\frac{1}{T_k}\Big] = -\frac{\alpha}{\mu_k(-\alpha-1)(L-\alpha)}.
	\end{equation}
	Then, by differentiating both sides the Expression \eqref{E:EVBL} with respect to the $L$, we have
	\begin{equation}\label{E:EVBL1}
		\E\Big[\frac{Z_n}{T_k^2}\Big] = -\frac{\alpha}{\mu_k(-\alpha-1)(L-\alpha)^2}.
	\end{equation}
	With this
	\begin{align*}
		\E(U_{{\bm \beta^\top} L}) &= {\bm 0}_{p+1}.
	\end{align*}

	To obtain $U_{\alpha \alpha}$, we employ the Expression \eqref{E:SF_A}.
	Thus, we have:
	$$
	U_{\alpha \alpha} = \frac{\partial U_{\alpha}}
	{\partial \alpha}
	= n U_1^{(1)}(\alpha,L)-2\sum_{k=1}^n\frac{\mu_k}{T_k}-(\alpha-L)\sum_{k=1}^n\frac{\mu_k^2}{T_k^2},
	$$
	where $U_1^{(1)}(\alpha,L)$ is the first derivate of $U_1(\alpha,L)$ with respect to the parameter $\alpha$.
	Applying the expected value in expression above yields:
	\begin{equation*}
		\E(U_{\alpha \alpha}) = n U_1^{(1)}(\alpha,L)-2\sum_{k=1}^n\mu_k\E\Big(\frac{1}{T_k}\Big)
		-(\alpha-L)\sum_{k=1}^{n}\mu_k^2 \E\Big(\frac{1}{T_k^2}\Big).
	\end{equation*}
	Using Corollary~\ref{C:EV_T} in this expression, we obtain:
	\begin{equation*}
		\E(U_{\alpha \alpha}) = n c_3,
	\end{equation*}
	where
	\begin{equation*}
		c_3 = U_1^{(1)}(\alpha,L)+\frac{2\alpha}{(L-\alpha)(-\alpha-1)}+
		\frac{\alpha(\alpha-1)}{(-\alpha-1)^2(L-\alpha+1)}.
	\end{equation*}

	From Expression \eqref{E:SF_A}, the second derivative of $\ell({\bm \theta})$ with respect to the $L$ can be written  as
	\begin{equation*}
		U_{\alpha L} = \frac{\partial U_{\alpha}}{\partial  L }
		= -n \Psi^{(1)}(L-\alpha)+\sum_{k=1}^n\frac{z_k}{T_k}+\sum_{k=1}^{n}\frac{\mu_k}{T_k}
		+(\alpha-L)\sum_{k=1}^n\mu_k\frac{z_k}{T_k^2},
	\end{equation*}
	where
	$\Psi^{(k)}(x)={\partial^{k+1}\log \Gamma(x)}/{\partial x^{k+1}}$ for $x > 0.$
	By applying the expected value, we obtain:
	\begin{multline}\label{E:UAL}
		\begin{split}
			\E(U_{\alpha L}) = -n\Psi^{(1)}(L-\alpha) + \sum_{k=1}^n\E\Big(\frac{Z_k}{T_k}\Big)
			+\sum_{k=1}^{n}\mu_k\E\left(\frac{1}{T_k}\right) \\
			+(\alpha-L)\sum_{k=1}^{n}\mu_k\E\Big(\frac{Z_k}{T_k^2}\Big).
		\end{split}
	\end{multline}

	Then, for expression
	$\E({Z_k}/{T_k})$ we invoke
	the fact that $\E(U_\alpha)=0$, that is,
	\begin{equation*}
		\sum_{k=1}^n\E(\log T_k) =
		-n U_1(\alpha,L)+\sum_{k=1}^n\log\mu_k+(\alpha-L)\sum_{k=1}^n\mu_k \E\Big(\frac{1}{T_k}\Big).
	\end{equation*}
	By differentiating both sides the expression above with respect the $ L $, we obtain
	\begin{align}\label{E:UA}
		\sum_{k=1}^{n}\E\left(\frac{Z_k}{T_k}\right) =
		n\Psi^{(1)}( L -\alpha) - \sum_{k=1}^{n}\mu_k\E\left(\frac{1}{T_k}\right)
		-(\alpha- L )\sum_{k=1}^{n}\mu_k \E\left(\frac{Z_k}{T_k^2}\right).
	\end{align}
	Hence, using the Expression \eqref{E:UA} in \eqref{E:UAL}, we have:
	$$
	\E(U_{\alpha L})= {0}.
	$$
	Ultimately, we have for $U_{L L}$ using the Expression
	\eqref{E:UL}, this is,
	\begin{align*}
		U_{LL}=\frac{\partial U_L}{\partial  L} =
		n \Big[\Psi^{(1)}(L-\alpha)-\Psi^{(1)}(L)+\frac{1}{L}\Big]
		- 2\sum_{k=1}^n\frac{z_k}{T_k}-(\alpha-L)\sum_{k=1}^n\Big(\frac{z_k}{T_k}\Big)^2.
	\end{align*}
	Applying the expected value yields:
	\begin{align*}
		\E(U_{LL}) =
		n \Big[\Psi^{(1)}(L-\alpha)-\Psi^{(1)}(L)+\frac{1}{L}\Big]
		-2\sum_{k=1}^n\E\Big(\frac{Z_k}{T_k}\Big)-(\alpha-L)\sum_{k=1}^n
		\E\Big(\frac{Z_k}{T_k}\Big)^2.
	\end{align*}
	Using~\eqref{E:UA} and the Expressions \eqref{E:EVBL} and \eqref{E:EVBL1}, we obtain:
	$$
	\sum_{k=1}^n\E\Big(\frac{Z_k}{T_k}\Big) = n \Psi^{(1)}(L-\alpha).
	$$
	By differentiating both sides of the expression above with respect the $L$, we get:
	$$
	\sum_{k=1}^n \E\Big(\frac{Z_k}{T_k}\Big)^2 = -n \Psi^{(2)}(L-\alpha).
	$$
	Then,
	$$
	\E(U_{LL}) = n c_4,
	$$
	where
	$$
	c_4 = (\alpha-L)\Psi^{(2)}(L-\alpha)-\Psi^{(1)}(L-\alpha)-\Psi^{(1)}(L)+\frac{1}{L}.
	$$
	Therefore it follows that the Fisher information matrix for ${\bm \theta}=(\alpha,{\bm \beta}^\top,L)^\top$ is given in \eqref{E:FIM}.

	We partition this matrix to obtain the inverse of $K({\bm \theta})$:
	$$
	{\bm K}({\bm \theta}) =
	\begin{bmatrix}
		{\bm A}      & {\bm B}
		\\[3mm]
		{\bm B}^\top & {\bm D}
	\end{bmatrix},
	$$
	where
	$$
	{\bm A} =
	\begin{bmatrix}
		\alpha{\bm X}^\top{\bm W}{\bm X} & c_2{\bm X}^\top\bm{E}{\bm \mu}^\ast
		\\
		c_2{\bm X}^\top\bm{E}{\bm \mu}^\ast                & nc_3
	\end{bmatrix},
	\quad
	{\bm B}^\top = {\bm B} =
	{\bm 0}_{p+2},
	$$
	and ${\bm D} = n c_4$.
	The above matrices can be blockwise inverted (see, e.g., \citet[p.~33]{Rao1973} and  \citet{Rencher2007}), therefore, we have that:
	$$
	{\bm K}^{-1}({\bm \theta})
	=
	\Bigg(\begin{array}{cc}
		{\bm A}^{-1} + {\bm \upsilon}{\bm \Phi}^{-1}{\bm \upsilon}^\top &
		- {\bm \upsilon}{\bm \Phi}^{-1} \\ \\
		-{\bm  \Phi}^{-1}{\bm \upsilon}^\top  &   {\bm \Phi}^{-1}         \\
	\end{array}\Bigg),
	$$
	where
	${\bm  \Phi} = {\bm D} - {\bm B}^\top{\bm A}^{-1}{\bm B}$,
	${\bm \upsilon} = {\bm A}^{-1}{\bm B}$.
	As ${\bm B}^\top={\bm B} = [{\bm 0}{\bm 0}]$, we have:
	$$
	{\bm K}^{-1}({\bm \theta}) =
	\begin{pmatrix}
		{\bm A}^{-1} & {\bm 0}_{p+2}\\
		{\bm 0}_{p+2}^\top     & (nc_4)^{-1}
	\end{pmatrix},
	$$
	with
	$$
	{\bm A}^{-1}
	=
	\Bigg(\begin{array}{cc}
		(\alpha{\bm X}^\top{\bm W}{\bm X})^{-1} + {\bm \zeta}
		{\bm \vartheta}^{-1}{\bm \zeta}^\top &
		-{\bm \zeta}{\bm \vartheta}^{-1} \\
		-{\bm \vartheta}^{-1}{\bm \zeta}^\top & {\bm \vartheta}^{-1}
	\end{array}\Bigg),
	$$
	where
	$${\bm \vartheta} = n c_3 - \frac{c_2^2}{\alpha}
	({\bm X}^\top\bm{E}{\bm \mu}^\ast)^\top
	({\bm X}^\top{\bm W}{\bm X})^{-1}
	({\bm X}^\top\bm{E}{\bm \mu}^\ast),$$
	and
	$$
	{\bm \zeta} = \frac{c_2}{\alpha}
	({\bm X}^\top{\bm W}{\bm X})^{-1}
	({\bm X}^\top\bm{E}{\bm \mu}^\ast).
	$$

	\section{Diagnostic measures}\label{App5}

	In this appendix, we derive the generalized leverage for $({\bm \beta}^\top,\alpha,L)$.
	Thus, from \eqref{E:SF_B}, we have
	$$
	D_{\bm \beta} = \sum_{k=1}^n
	\frac{\partial \mu_k}{\partial{\bm \beta}} =
	\sum_{k=1}^n\frac{1}{g^\prime(\mu_k)}\bm{x}_{k} =
	\bm{E}{\bm X}.
	$$
	From Equation~\eqref{E:DBB}, it follows that
	\begin{align*}
		\Big[-\frac{\partial^2 \ell({\bm \theta})}
		{\partial {\bm \beta} \partial {\bm \beta}^\top}\Big]_{jl} =
		\sum_{k=1}^n
		\bigg\{
		\bigg[
		\frac{\alpha}{\mu_k^2}+\frac{c_1(-\alpha-1)}{T_k^2}
		\bigg]
		+\Big[
		\frac{\alpha}{\mu_k} + \frac{c_1}{T_k}
		\Big]\frac{g^{\prime\prime}(\mu_k)}{g^{\prime}(\mu_k)}
		\bigg\}
		\frac{1}{g^\prime(\mu_k)^2} x_{k j}x_{k l},
	\end{align*}
	and
	$$
	-\frac{\partial^2 \ell({\bm \theta})}
	{\partial {\bm \beta} \partial {\bm \beta}^\top}= \alpha{\bm X}^\top{\bm Q}{\bm X}.
	$$
	Also, it can be shown that: The $k$th column of the matrix ${\partial^2 \ell({\bm \theta})}/
	{\partial {\bm \beta} \partial {\bm z}^\top}$ is
	\begin{align*}
		\frac{\partial \mu_k}{\partial g(\mu_k)}
		\frac{\partial g(\mu_k)}{\partial {\bm\beta}}
		\frac{\partial^2 \ell({\bm \theta})}{\partial \mu_k\partial { z_k}}
		= \alpha
		\frac{{\bm x}_k}{g^\prime(\mu_k)}
		\frac{ L }{T_k^2}\frac{c_1}{\alpha}
	\end{align*}
	and, therefore, its matrix form is
	$$
	\frac{\partial^2 \ell({\bm \theta})}
	{\partial {\bm \beta} \partial {\bm z}^\top}
	=
	\alpha{\bm X}^\top\bm{E}{\bm T^\ast}.
	$$

	\section*{Acknowledgements}

	Authors would like to thank the National Council for Scientific and Technological Development, Grant/Award Number: CNPq 200819/2022-4.

\onecolumn

{\small
\singlespacing

}

\end{document}